\documentclass[twocolumn,a4paper]{aa}
\usepackage{graphicx}
\usepackage{amsmath}
\usepackage{wrapfig}
\usepackage[utf8]{inputenc}
\usepackage{float}
\usepackage{dblfloatfix}

\usepackage{breqn} 

\newcommand{\electron}{el}

\newcommand{\T}{{\Large$\mathstrut$}}
\newcommand{\TT}{{\LARGE$\mathstrut$}}

\begin{document}
  \title{The shape of the Photon Transfer Curve of CCD sensors}
\author{
  Pierre~Astier\inst{\ref{aff:LPNHE-CNRS}}
  \and
  Pierre~Antilogus\inst{\ref{aff:LPNHE-CNRS}}
  \and
  Claire~Juramy\inst{\ref{aff:LPNHE-CNRS}}
  \and
  Rémy~Le~Breton\inst{\ref{aff:LPNHE-SU}}
  \and
  Laurent~Le~Guillou\inst{\ref{aff:LPNHE-SU}}
  \and
  Eduardo~Sepulveda\inst{\ref{aff:LPNHE-CNRS}}
  \and
  \\ The Dark Energy Science Collaboration
}
\institute{%
  {LPNHE, (CNRS/IN2P3, Sorbonne Université, Paris Diderot), 
  Laboratoire de Physique Nucléaire et de Hautes Énergies,
  F-75005, Paris, France}\label{aff:LPNHE-CNRS}
  \and
  {Sorbonne Université, Paris Diderot, CNRS, IN2P3,
  Laboratoire de Physique Nucléaire et de Hautes Énergies, LPNHE,
  F-75005, Paris, France}\label{aff:LPNHE-SU}
}
\titlerunning{PTC shape}
\offprints{pierre.astier@in2p3.fr}
\date{Received Mont DD, YYYY; accepted Mont DD, YYYY}
\abstract{
The Photon Transfer Curve (PTC) of a CCD depicts the variance
of uniform images as a function of their average. It is now well
established that the variance is not proportional to the average, as
Poisson statistics would indicate, but rather flattens out at high
flux. This ``variance deficit'', related to the brighter-fatter effect,
feeds correlations between nearby pixels, that increase with
flux, and decay with distance. We propose an analytical expression for
the PTC shape, and for the dependence of correlations with intensity,
and relate both to some more basic quantities related to the
electrostatics of the sensor, that are commonly used to correct
science images for the brighter-fatter effect. We derive electrostatic
constraints from a large set of flat field images acquired with a
CCD e2v 250, and eventually 
question the generally-admitted assumption that boundaries of CCD pixels 
shift by amounts proportional to the source charges. Our results
  show that the departure of flat field statistics from Poisson law is entirely compatible with charge redistribution during the drift in the sensor.
} 

\maketitle

\section{Introduction}
The response of CCD sensors to uniform illumination is used
primarily to probe the sensor cosmetics (and possibly correct
science images for some defects), to compensate for possible response
non uniformity, and to measure the gain of the 
sensor and its electronic chain up to the digitization.
For astronomy, the gain of each video channel (expressed 
in \electron/ADU) allows one to evaluate the expected Poisson
variance of a pixel from its value, which is required to
propagate the shot noise to flux and position measurements
performed on astronomical images.

Since Poisson fluctuations drive the variance of uniform illuminations (commonly called flat fields), the gain could easily
be obtained as the ratio of the average to the variance of a flat field. (Note that the gain used
by astronomers varies in the opposite way as the electronic gain).
In order to allow for a constant read out noise (very
sub-dominant in practice), one usually performs a series of
flat field images at increasing illuminations in order to separate the two
contributions. This variance versus average of uniform illuminations is
called the Photon Transfer Curve (PTC) of a sensor (or more
appropriately of a video channel), and was first introduced,
  as a mean to measure gain and read noise, in \cite{JanesickPTC85}.
When the read out noise becomes
negligible, the variance should just increase as the mean, if
it follows Poisson statistics. In \cite{Downing06}, it is shown
that the PTC of CCD sensors is not linear but rather flattens out at
high fluxes: close to saturation of the sensor, one can miss up to 25~\%
of the variance expected from extrapolating the slope at low flux. The
same authors remark that re-binning the image (i.e., summing nearby
pixels into bigger ones) improves very efficiently the linearity of the
PTC curve. One can readily infer from these observations that nearby
pixels of the original images are positively correlated (at least on
average), which is indeed reported in \cite{Downing06}.  One can also
deduce that the covariances between nearby pixels in uniform
illuminations grow faster than their average and hence that
correlations grow with the average: indeed the correlations shown in
\cite{Downing06} are compatible with a linear increase with the
average, and hence the variance should be a quadratic function of the
average. One should note that even if the electronic chain is
perfectly linear, this is a genuine departure from linearity, because adding
two uniform exposures of (e.g.) 10~ke on average does not give the
variance expected in a 20 ke exposure. So, in a 20~ke exposure, the
second half is sensitive to the fluctuations in the first half.
It is then in principle possible to distinguish a 20~ke flat field exposure 
from the sum of two 10~ke exposures. 

In flat fields, this non-linearity is only noticeable on variance and
covariances, but in structured images, as we will discuss shortly, the
average {\it values} do not add up. Around 2012, at least three teams
noticed that stars on astronomical images, or spots on CCD test
benches tend to become slightly bigger when they become brighter (see
\citealt{Antilogus14, Lupton14}, App. B of \citealt{Astier-photom-13}).
This broadening is nowadays refered to as the "brighter-fatter effect".
It was probably noticed at these times due to the advent of large 
observational programs relying on thick fully-depleted 
sensors (for their high NIR efficiency) which make 
the effect bigger than in thinned CCDs \citep{Stubbs14}. 

In \cite{Antilogus14, Guyonnet15}, it is proposed that this departure
from linearity is due to the electric fields sourced by the charges
stored in the pixel potential wells, which increase as image integration
goes. Pixels containing more charges than their neighbors will tend to
repel forthcoming charges, hence reducing their own effective physical
area. One can derive the size of the expected effects from electrostatics, 
and the expected effect size is shown to be broadly compatible
with the observations in \cite{Antilogus14}, and much more precisely
in \cite{Lage17}.

The brighter-fatter effect could bias the PSF size by  $\sim$
  1\% or more for faint sources, which would bias PSF fluxes of
  faint point sources by the same amount; this is no longer tolerable
  for e.g. supernova Ia cosmology \citep[see e.g.,][]{Betoule14}.  It
  would also bias the shear of faint galaxies by even more, which is
  even less acceptable \citep[for
    example][]{Jarvis-PACCD-14,Mandelbaum-PACCD-14}. In both cases,
  biases on this scale are just not acceptable for current large
  imaging surveys, and in this context, one {\it has} to devise a
  precise correction.  So far, all attempts to correct for the
brighter-fatter effect have relied on the pixel correlation function
in flat fields to infer the modifications to pixel boundaries sourced
by a given astronomical scene \citep{Gruen-PACCD-15,Coulton-17}, an
approach proposed in \cite{Antilogus14, Guyonnet15}. LSST\footnote{For a general presentation of LSST, see
    e.g. \cite{IvezicLSST08} or visit \texttt{http://lsst.org.}}
  strategy to handle the effect follows the same lines, and still
  anticipate to use flat field statistics to infer the alteration
  of pixel areas caused by stored charges. This paper, as compared to
  previous enterprises, refines the relation between flat field
  statistics and pixel area alterations. We also discuss several
  technicalities of covariance measurements, on a practical example,
  and show that currently used simplifications can cause very
  significant biases of the pixel area alterations and hence of the
  brighter-fatter effect correction. We do not discuss here the
  practical correction of astronomical scenes, described in the above
  references, noting however that they all assume that images are
  sufficiently well sampled to infer the incoming charge flow over
  pixel boundaries.

In this paper we first evaluate how variances and covariances
in flat fields grow with their average (\S \ref{sec:PTC_shape},
\S \ref{sec:cov_dynamics}), in order to infer
as precisely as possible the strength of electrostatic interactions
required to constrain any empirical electrostatic model, eventually
used to ``undo'' the brighter-fatter effect. These measurements
  are necessary whether one uses a mostly  agnostic approach as in
  \cite{Guyonnet15, Gruen-PACCD-15, Coulton-17} or a numerical solution
  of Poisson equation as in \cite{Lage17}.
Section \ref{sec:quest_linearity} generalizes the approach
to pixel boundary shifts not exactly proportional to the source charges. 
We describe the laboratory setup used to generate data (\S \ref{sec:meas}),
and describe our analysis of a large flat field image set
and our findings in \S \ref{sec:analysis}. We summarize and conclude
in \S \ref{sec:discussion}.

\section{The importance of PTC and covariance curves shapes}
\label{sec:PTC_shape}
The flattening of the PTC can be described trivially by 
adding one order to the polynomial used to fit the curve, i.e., replacing
a linear fit (expected if Poisson statistics hold, with some read noise contribution) by a quadratic one. This is somehow justified by the fact
that if one models the pixel covariances induced by electrostatics,
they scale, to first order, as the product $V\mu$ (variance times average) of flat fields
\citep{Antilogus14}.
Since $V$ scales as $\mu$ to zeroth order, and the total noise power 
(variance plus covariances) is conserved by the electrostatic re-distribution,
this indicates that the sum of covariances takes away a quadratic
contribution ($\propto \mu^2$) from the variance as flux increases.
This paradigm roughly describes the data, so why should we worry
about ``higher orders'' or ``next to leading order'' (NLO) effects? To
set the scale, the ``variance deficit''(i.e. by how much the variance
of flat fields is lower than the Poisson expectation) can reach 20\% at
100~ke intensity. It indicates that higher order corrections could be of
the order of 4\% at the same intensity. As ignoring the flattening of
the PTC biases gain estimations, ignoring the same effect on
correlations biases the measurement, possibly by a similar amount. For
example, ignoring the quadratic behavior of the PTC would typically bias the
gain by 10\% if the variance deficit is 20 \% at the high end.

While it is easy to measure the second order on a PTC, it becomes more
involved for covariances, because their uncertainties are of the same
order as for PTC, but the values are at least 2 orders of magnitude
lower. This is why we work with data sets of typically 1000 flat field pairs
or more, while the shape of the PTC can be characterized with less than
100 pairs.

A model for the PTC shape is proposed in \cite{Rasmussen16}. It relies
on a simplified electrostatic model of the sensor to derive a PTC
shape model, which describes the data better than a parabola, but still
relies on approximations that we will avoid. The electrostatics worked
out there indicates that assuming that pixel area distortions scale as
source charges is probably not entirely true, as the vast majority of
previous works assume. This work also tackles the contributions of
diffusion to the brighter-fatter effect and confirms that they are
largely sub-dominant.

We now establish the relation between some electrostatics-related quantities
and the shape of PTC
and covariance curves, beyond perturbative arguments.

\section{Dynamical development of variance and covariances}
\label{sec:cov_dynamics}
We aim here at modeling the time-dependent build-up of correlations in
flat fields. We will first express the basic (differential) equations
governing the phenomenon, and then solve them.

As common in electrostatics, we distinguish source charges 
from test charges. When describing a flat field filling up,
the source charges are the ones stored 
in the potential wells of the sensor, and the test charges are the ones
drifting in the sensor. 
We label `` $00$'' a particular pixel (far from the edges of the sensor),
and index its neighbors by their coordinates with respect to this ``central''
pixel: $Q_{ij}$ refers to a pixel located $i$ columns and $j$ rows away from
the pixel $(0,0)$. The charge $Q_{ij}$ alters the current impinging on pixel $(0,0)$, by modifying the drift lines and consequently shifting the pixel boundaries.
The current flowing into pixel $(0,0)$ reads:
\begin{equation}
  {\dot Q_{00}} = I [ 1 + \sum_{kl} a_{kl} Q_{kl}]
  \label{eq:interaction}
\end{equation}
where $a_{kl}$ describes the strength (and sense) of the interaction,
and $I$ is the current that would flow in the absence of interactions (all $a_{kl} = 0$). Since we are considering uniform exposures, $I$ does not vary with position, nor time. 
The coefficients $a_{kl}$
describe the change of pixel area per unit stored charge caused at a pixel
located at a separation $(k,l)$ from the source, where $k$ is the separation along rows (the serial direction), and $l$ the separation along columns (i.e. the parallel direction). In \citealt{Antilogus14}, coefficients labeled as
$a^X_{ij}$ describe the pixel boundary shifts induced by stored charges.
The relation between boundary shifts and change of area can be expressed
(to first order) as 
$a_{ij} = \sum_X a^X_{ij}$, where the sum runs over the four sides
of a pixel.

The equation \ref{eq:interaction} relies on the fact that
electrostatic forces are proportional to source charges $Q_{kl}$, and assumes
that the alteration of pixel area is proportional to the charge that is
causing it. The latter is not a prescription of electrostatics.
Thanks to parity symmetry, these $a_{kl}$ coefficients only depend on the absolute
value of $k$ and $l$. If we consider a single source charge $Q_{kl}$, the sum of currents flowing into all affected pixels should not depend on this source charge. This imposes the following sum rule :
\begin{equation}
  \sum_{kl} a_{kl}   = 0 \label{eq:sum_rule}
\end{equation}
where the sum runs over positive, null and negative $k$ and $l$.
Since $a_{00}$ 
describes the change of a pixel area due to its own charge content,
and since same-charge carriers repel each other, this pixel area has to shrink
as charge accumulates inside the pixel, which implies $a_{00}<0$.
Since the $a_{ij}$ are almost
always positive, the sum rule imposes that $a_{00}$ is much larger in
absolute value than any other $a_{ij}$. Since the sum in eq. \ref{eq:sum_rule}
necessarily converges, $a_{ij}$ should decay faster than $r^{-2}$ with
$r \equiv \sqrt{i^2+j^2}$.

In order to describe the shape of the variance versus average curve,
and the associated covariances, we evaluate $Cov[{\dot Q_{00}}, Q_{ij}]$.
We will handle $(i,j) = (0,0)$ later and first concentrate
on $(i,j) \neq (0,0)$. In this latter case, only the second term of eq.
\ref{eq:interaction} matters:
\begin{align}
  Cov[{\dot Q_{00}}, Q_{ij}] &= I \sum_{kl} a_{kl} \ Cov(Q_{kl}, Q_{ij}) \nonumber \\
  &=  I \sum_{kl} a_{kl} C_{i-k, j-l} 
\end{align}
where $C_{ij}$ denotes the covariance of pixels located at separation $(i,j)$, hence $C_{00}$ is the variance. We then have (still for $(i,j) \neq (0,0)$):
\begin{equation}
  {\dot C}_{ij} =  Cov[{\dot Q_{00}}, Q_{ij}] +  Cov[Q_{00}, {\dot Q_{ij}}] = 2 I \sum_{kl} a_{kl} C_{i-k, j-l} 
\end{equation}
where the two covariances are equal because of parity symmetry.

When $(i,j) = (0,0)$ there is an extra contribution coming from
$Cov(I_{00}, Q_{00})$ where $I_{00}$ refers to the current flowing
into the undistorted pixel $(0,0)$, with its statistical fluctuations. For a
Poisson process, this reads $V_I/2$ where $V_I$ is the average
number of quanta per unit time in the current $I_{00}$.
If all $a_{kl}$ are zero, we get ${\dot C}_{00} = V_I$, and $C_{00}(t)=V_I t$,
which is just the Poisson variance.

So, we rewrite the above equation for all $(i,j)$ values as:
\begin{equation}
  {\dot C}_{ij} = \delta_{i0} \delta_{j0} V_I + 2 I \sum_{kl} a_{kl} C_{i-k, j-l} \label{eq:diff_eq}
\end{equation}
The sum on the right-hand side contains a ``direct'' term $a_{ij}
C_{00}$ where the fluctuations of the charge at $(i,j)$ source the
covariance. But all covariances involving the source (on the RHS) and
any other pixel also feed covariances (on the LHS). Of course, these
``three-pixel terms'' are expected to be small, but they are numerous,
and we should track them down in the analysis.

If we sum eq. \ref{eq:diff_eq} over all separations, we have:
\begin{align}
  \sum_{ij}   {\dot C}_{ij} & = V_I + 2 I \sum_{ij} \sum_{kl} a_{kl} C_{i-k, j-l} \nonumber \\
  & = V_I + 2 I \sum_{kl}  a_{kl} \sum_{ij} C_{i-k, j-l} \label{eq:sum_conv1} \\
  & = V_I + 2 I \sum_{kl} a_{kl} \sum_{ij} C_{ij} \label{eq:sum_conv2} \\
  &= V_I \label{eq:sum_dcov_dt}
\end{align}
where the last step follows from the sum rule $\sum_{ij} a_{ij} = 0$,
and one goes from equation \ref{eq:sum_conv1} to \ref{eq:sum_conv2}
by noting that $\sum_{ij} C_{i-k, j-l}$ does not depend on $k$ or $l$.
We then have $\sum_{ij} C_{ij} = V_It$, i.e. the sum of variance and
covariances is the Poisson variance $V_I t$. This indicates that if
one rebins the uniform image into big enough pixels, the Poisson behavior
of the variance vs average curve is restored \citep[as originally
 reported in][]{Downing06}.  Alternatively, imposing that $\sum_{ij} C_{ij}
= V_It$ yields $ \sum_{ij} a_{ij} = 0$. Note that the derivation above does
not assume that $a_{ij}$ is independent of time (i.e. accumulated charge).

One can rewrite the differential equation \ref{eq:diff_eq} as:
\begin{equation}
  \dot{\boldsymbol{C}} = \boldsymbol{\delta} V_I + 2I \boldsymbol{C} \otimes \boldsymbol{a} \label{eq:C_conv}
\end{equation}
where $\boldsymbol{C}$ refers to the 2-d array of covariances, and $\boldsymbol{\delta}$ to the
2-d discrete delta function, and the symbol $\otimes$ refers to discrete convolution. One can solve for $\boldsymbol{C}$ as a series of powers of $t$,
but it is tempting to apply a discrete Fourier transform in pixel
space (i.e. over the spatial indices
involved in the convolution) to this differential equation, so that the convolution product becomes
a regular product. The equation system then becomes diagonal:
\begin{equation}
  \tilde{\dot{\boldsymbol{C}}} = V_I + 2 I \tilde{\boldsymbol{a}} \tilde{\boldsymbol{C}}
\end{equation}
where $\tilde{\boldsymbol{C}}$ refers to the (spatial) Fourier transform of $\boldsymbol{C}$,
and similarly for $\boldsymbol{a}$.
These equations are now independent for each ``separation'' in reciprocal space.
We assume that the $\boldsymbol{a}$ coefficients are constants, i.e. are independent
of time or accumulated charge;
we impose  $\tilde{\boldsymbol{C}}(t=0) = 0$, and the solution reads:
\begin{equation}
  \tilde{\boldsymbol{C}}(t)= \frac{V_I}{2I\tilde{\boldsymbol{a}}}\left[ e^{2 I \tilde{\boldsymbol{a}} t } -1 \right ] 
\end{equation}
And the Taylor expansion\footnote{Relying on a series expansion is
  mathematically justified, because the exponential series has an
  infinite radius of convergence. From a more practical point of view,
  the products $aIt$ are small (at most $\sim 0.2$), and hence the size of
  the successive terms decays rapidly.} reads:
\begin{align}
  \tilde{\boldsymbol{C}}(t) &= V_It \left[1 + I\tilde{\boldsymbol{a}}t + \frac{2}{3} (I\tilde{\boldsymbol{a}}t)^2 + \frac{1}{3}(I\tilde{\boldsymbol{a}}t)^3 + \cdots \right]  \\
  \tilde{\boldsymbol{C}}(\mu) &= V \left[1 + \tilde{\boldsymbol{a}} \mu + \frac{2}{3} (\tilde{\boldsymbol{a}} \mu )^2 + \frac{1}{3}(\tilde{\boldsymbol{a}}\mu)^3 + \cdots \right]\nonumber
\end{align}
where $V \equiv V_It$ is the Poisson variance of the image (i.e. the variance for $\boldsymbol{a} = 0$), and $\mu \equiv It$ is its average (unaffected by $\boldsymbol{a}$, because charge is conserved). Transforming back to direct pixel space, we obtain:
\begin{equation}
  \boldsymbol{C}(\mu) = V \left[\boldsymbol{\delta} + \boldsymbol{a} \mu + \frac{2}{3} TF^{-1}[(\tilde{\boldsymbol{a}})^2] \mu^2 + \cdots \right] 
\end{equation}
where $\boldsymbol{\delta}$ refers again to the 2-d discrete delta function in pixel space,
and $TF^{-1}$ to the inverse Fourier transform. Since $V$ is the Poisson variance,
it is proportional to $\mu$ and we use the common definition of ``gain''
used in astronomy $V=\mu/g$ and further add a constant term to describe the
contribution of electronic noise (and its correlations). Making the components explicit,
we get:
 \begin{dmath} 
   C_{ij}(\mu) =  \frac{\mu}{g} \left[\delta_{i0} \delta_{j0} + a_{ij} \mu + \frac{2}{3} [\boldsymbol{a} \otimes \boldsymbol{a}]_{ij} \mu^2 \\
     + \frac{1}{3}[\boldsymbol{a}\otimes \boldsymbol{a} \otimes \boldsymbol{a}]_{ij} \mu^3 + \cdots \right] + n_{ij}/g^2\label{eq:C_ij_adu}
 \end{dmath}
By expressing the constant term as $n_{ij}/g^2$, $n$ is defined in el$^2$ units.

The analysis presented in \cite{Antilogus14,Guyonnet15} roughly
correspond to the two first terms in the bracket: the variance
$C_{00}$ is approximated by a parabola, and the correlations ($C_{ij}/V$) are
linear in $\mu$ (and $a_{ij}$ is their slope, commonly positive). At
this level of approximation, there is a trivial relation between the
values of $\boldsymbol{a}$ and the measured covariances. Higher order terms mix the
relation between the measured $C_{ij}$ and the interaction strengths
$a_{ij}$.  Since $a_{00}<0$ for all types of CCDs, the variance of flat fields grows
less rapidly than their average.

Since the $\boldsymbol{a}$ quantities carry an inverse charge unit, their values are sensitive to the charge
unit. If expressing charges in ADU is straightforward, it makes $\boldsymbol{a}$
dependent on the electronic gain, which seems inadequate for quantities
attached to the sensor itself. So, we decide to express $\boldsymbol{a}$ in inverse \electron,
and expression \ref{eq:C_ij_adu} becomes:
\begin{dmath} 
  C_{ij}(\mu) =  \frac{\mu}{g} \left[\delta_{i0} \delta_{j0} + a_{ij} \mu g + \frac{2}{3} [\boldsymbol{a} \otimes \boldsymbol{a}]_{ij} (\mu g)^2
    + \frac{1}{3}[\boldsymbol{a} \otimes \boldsymbol{a} \otimes \boldsymbol{a}]_{ij} (\mu g)^3 + \cdots \right] + n_{ij}/g^2 \label{eq:C_ij}
\end{dmath}
where both $C_{ij}$ and $\mu$ are expressed in ADU, i.e. as measured.
The generic numerical factor of the term $(\mu g)^n$ reads $2^n/(n+1)!$. Regarding the expansion truncation, we reproduced the analysis that follows with one extra order, and the results are almost indistinguishable. However, for some separations, adding this extra term changes the predictions by as much as 2\% for $\mu$ close to saturation.

One may note that the conservation of variance
  (i.e. $\sum_{ij} C_{ij} = \mu/g$, as discussed earlier, see eq. \ref{eq:sum_dcov_dt}) is ensured because when summed
  over all separations, all terms but the first in the RHS bracket of the
  above equation vanish. This is a consequence of the sum rule
  $\sum_{ij} a_{ij} = 0$, because $\sum_{ij} [\boldsymbol{a} \otimes
    \boldsymbol{a}]_{ij} = [\sum_{ij} a_{ij}]^2$, for the same reason
  as the one used earlier between equations \ref{eq:sum_conv1} and
  \ref{eq:sum_conv2}. By recurrence, higher convolution powers of
  $\boldsymbol{a}$ also integrate to $0$.

For the shape of the PTC, one can think to approximate the values of
convolutions as $a_{00}^n$ since $a_{00}$ dominates in size. This is a simple
way of approximating the PTC shape as a function of only one parameter,
but testing the validity of the approximation still requires to measure
the other $a_{ij}$ values. Within this approximation, the PTC shape reads:
\begin{equation}
C_{00} = \frac{1}{2 g^2 a_{00}} \left[ \exp \left(2 a_{00} \mu g \right) - 1 \right]+n_{00}/g^2
\end{equation}
where $a_{00}$ is negative.

One should remark that in equation \ref{eq:C_ij}, all terms of the
expansion are determined by the first order ($a\mu$). Neglecting
the term scaling as $a^2$ biases $a$ by a relative amount of order
$a\mu$, which reaches about 20\% at $\mu = 10^5$ \electron, for the sensor
we characterize later in the paper. So,
accounting for the term in  $a^2$ is mandatory when measuring $a$,
and in what follows, we have coded all the terms displayed in eq. \ref{eq:C_ij}.

In order to verify the above algebra, we have implemented a
Monte-Carlo simulation of eq. \ref{eq:interaction}, rewritten as:
\begin{equation}
  {\dot {\mathbf Q}} = I [ 1 + \boldsymbol{Q} \otimes \boldsymbol{a}]
\end{equation}
where ${\mathbf Q}$ is the pixelized image. The integration step reads:
\begin{equation}
  {\mathbf Q}(T_{i+1}) = {\mathbf Q}(T_i) + \mathrm{Poisson}\left[(T_{i+1} - T_i) I \left[ 1 + {\mathbf Q} \otimes \boldsymbol{a} \right] \right] \label{eq:mc_step}
\end{equation}
where ``Poisson'' refers to a Monte-Carlo realization of a Poisson law
having its argument as average (we used \verb|numpy.random.poisson|).
We have reduced the time step until results became stable to a few
$10^{-4}$ level and settled for $(T_{i+1} - T_i) I = 50$~\electron. We
have chosen $\boldsymbol{a}$ from an electrostatic simulation closely
describing our real data, integrated up to $\mu = 10^5$ el, and
generated numerous image sequences.  One integration step takes about
2s (on a core i7 CPU) for a 2k$\times$2k image.  We have then checked
that fitting eq. \ref{eq:C_ij} on covariances measured on the
simulated images delivers the input $\boldsymbol{a}$ to an acceptable
accuracy, for statistics similar to our real data set. This
  test on simulated data was mostly meant to confirm the algebra
  developed above and test the data reduction and fitting codes,
  because equation \ref{eq:mc_step} can be transformed into
  equation \ref{eq:C_conv} for a time step going to zero. One should then not
  expect subtle physical effects to show up in this simplified
  simulation.

\section{Questioning the linearity of the interaction model}
\label{sec:quest_linearity}
Our ``interaction equation'' (eq. \ref{eq:interaction}) assumes that
the strength of the current alteration is strictly proportional to the
source charge. This is questionable because both the shape and the
position of the charge cloud within the potential well may evolve as
charge accumulates. For example, a quantitative analysis
of such an effect can be found in \S 4.3 of \citealt{Rasmussen16}.
In the context of flat fields, one should expect
the drift field to go down as charge accumulates in the potential
wells. One can also consider the possibility that if the charge cloud
stored in a pixel well moves away from the parallel
clock stripes as charge flows in, electrostatic forces will increase
faster than the stored charges. One could think that covariances will
then grow faster than in the linear hypothesis (i.e. forces are just proportional to signal level), but covariances and
variance obey a sum rule, so anticipating the sense of the effect is
not straightforward. In any case, if such phenomena happen,
we expect that both the PTC and the covariances shapes are altered.

In order to quantitatively question this linear assumption, 
we propose to generalize eq. \ref{eq:interaction} by:
\begin{equation}
{\dot Q_{00}} = I [ 1 + \sum_{kl} a_{kl} Q_{kl} (1+b_{kl}\ I\ t)  ] \label{eq:interaction_mod}
\end{equation}
where we simply assume that the electrostatic force has an extra
component (presumably small) proportional to the square of the source
charge, as an extra term in a Taylor expansion. Charge conservation imposes a second sum rule : $\sum_{ij} a_{ij}b_{ij} = 0$.
We neglect the
contributions of random fluctuations but concentrate on the
average, and hence replace charges by their average when multiplied by
$\boldsymbol{b}$.  Note that $\boldsymbol{b}$ has the dimension of an inverse charge, as
$\boldsymbol{a}$.  We could not find an analytical expression for the solution, and
hence resorted to solving for a series in powers of $t$. Equation 
\ref{eq:C_ij} becomes:
\begin{dmath} 
  C_{ij}(\mu) =   \frac{\mu}{g} \left[\delta_{i0}\delta_{j0} + a_{ij} \mu g + \frac{2}{3} [\boldsymbol{a} \otimes \boldsymbol{a} + \boldsymbol{ab}]_{ij} (\mu g)^2 \\
    + \frac{1}{6}(2 \boldsymbol{a}\otimes \boldsymbol{a} \otimes \boldsymbol{a} + 5 \boldsymbol{a} \otimes \boldsymbol{ab})_{ij} (\mu g)^3 + \cdots \right] + \frac{n_{ij}}{g^2}\label{eq:C_ij_mod}
\end{dmath}
One can note that $\boldsymbol{b}$ only appears in the expression through the $\boldsymbol{ab}$ combination, as in the source equation \ref{eq:interaction_mod}.
Non-linearity of the interaction is mostly detected by the term in $\mu^2$ in the bracket being inconsistent with the linear term for $\boldsymbol{b}=0$. $\boldsymbol{a}$ and $\boldsymbol{b}$ are expressed in the same unit, i.e. inverse \electron, for $C_{ij}$ and $\mu$ expressed in ADU.

\section{Measurements}
\label{sec:meas}
We report here measurements performed on a CCD 250 from e2v that
has been developed for the LSST project (\cite{JuramySPIE2014,OConnorSPIE2016}). This
sensor is made of high-resistivity silicon, 100~$\mu$m thick, has 10~$\mu$m pixel side
and 4096x4004 pixels in total. It is divided into 16 segments, each
of which is 512x2002 pixels in size and has its own readout channel.
Each segment has its own serial register made of 522 pixels:
  it has 10 extra pre-scan pixels which are not fed by the science array.
  Following the vendor recommendations, we operate the sensor in full depletion mode.

\subsection{Laboratory setup}

\begin{figure*}[ht!]
\begin{center}
\includegraphics[width=\textwidth]{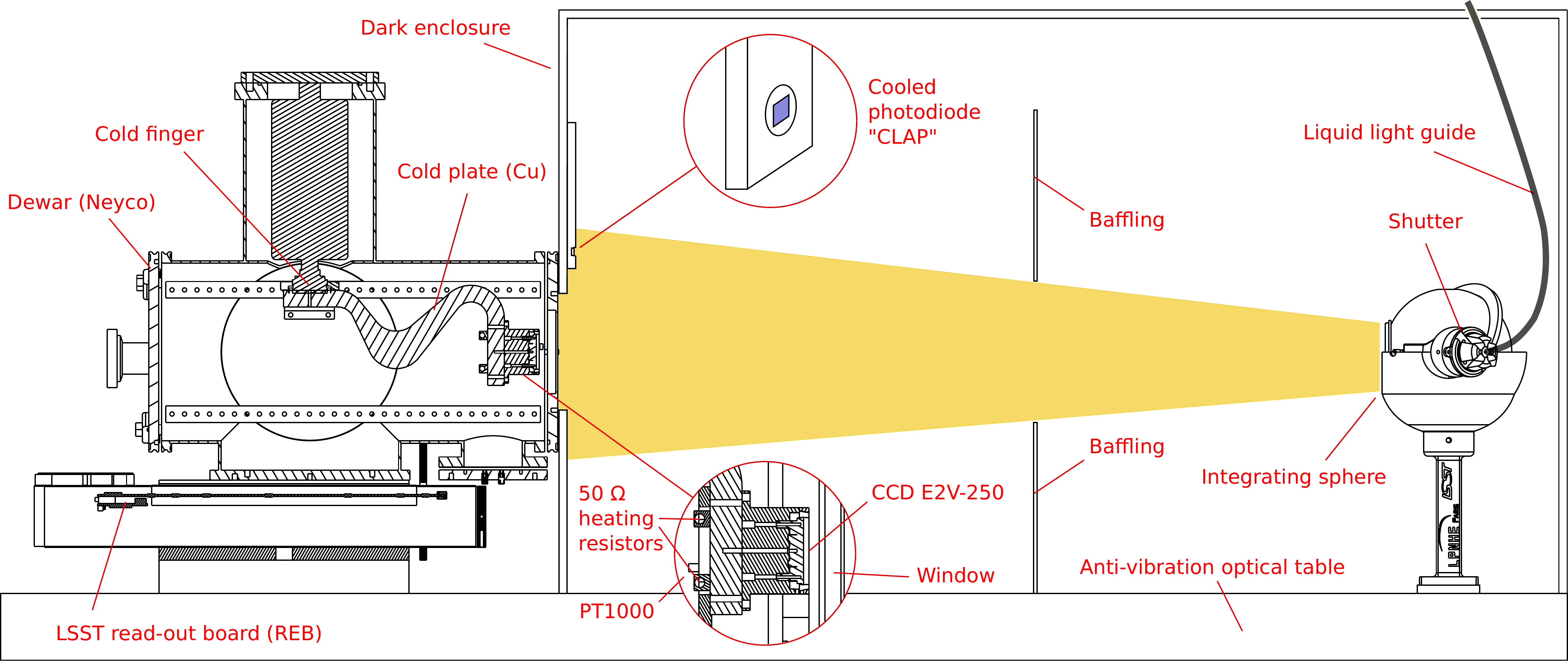}
\caption{Laboratory setup. The QTH lamp and the monochromator sit
  outside of the dark enclosure and are not drawn. The light from the
  monochromator exit slit is conveyed through a liquid light guide
  (Newport 77639 Liquid Light Guide, 2~m long, 8~mm diameter) which
  feeds the integrating sphere. Exposure time is controlled through a
  mechanical shutter placed between the end of the light guide and the
  sphere entrance port. Our cooled photodiode (``CLAP'') is attached
  on the dark box wall just above the dewar window, and allows to
  measure the effective exposure time (see fig.~\ref{fig:clap} below).
  \label{fig:bench}}
\end{center}
\end{figure*}

Our test CCD is temperature-controlled at $-100\pm0.01^{\circ}\text{C}$, is kept inside a Neyco Dewar at pressure below $7.10^{-7}$mbar, and is read out using the LSST electronic chain (see \cite{JuramySPIE2014,OConnorSPIE2016}), which runs at room
temperature on our test stand, in a dedicated class~10000 / ISO-7
clean room. The clean room temperature is regulated ($\pm 0.2^{\circ}$C)
to minimize temperature effects on the readout electronics.  
The video channels consist in a dual
slope integration (performed by two 8-channel ASICs named ASPIC, see \cite{JuramySPIE2014})
followed by a 18-bit ADC. In our setup, the CCD is connected to the
readout electronics by two flex cables (one per half CCD, and per
ASPIC), which also transport the CCD clocking lines. The sequencing of
read out is delivered by an FPGA driving the CCD clocks (through
appropriate power drivers) and the analog integration chain. We
typically read at 550~kpix/s (so that the image is read out in 2~s)
and measure a readout noise of $\sim$~5~el per pixel. The gains are
about 0.7~\electron/ADU. This setup achieved a gain stability of the full video chain over 3 days of a few $10^{-4}$ and always better than $10^{-4}$ within the 1-h time frame needed to measure the overall response
non-linearity once, as described below.

For flat field studies, our light source is a Newport 69931 QTH
(Quartz Tungsten Halogen) lamp, operated at a regulated power of
240~W, which feeds a monochromator via a lens and an order blocking
filter, when needed. For the data presented below, the monochromator
is set to 650~nm with a slit width of about 15~nm. The light is conveyed into
the dark box via an optical fiber (Newport 77639 Liquid Light Guide,
2~m long, 8~mm diameter). A mechanical shutter is placed between the end
of the fiber and the entrance port of the integrating sphere. A cooled
photodiode (Cooled Large Area Photodiode, ``CLAP'',
see~\cite{2015DICEpaper}, appendix~F) is attached to the dark box wall
and placed above the dewar window; it is readout using a low-noise
ASIC pre-amplifier, with $\sim$~300~Hz bandwidth, feeding a 31.25~kHz
flashing ADC. The Dewar is attached to one side of the dark box, and
sees the integrating sphere at a distance of $\sim$~1~m (see fig.~\ref{fig:bench}). The
illumination system delivers about 10,000~photoelectrons per second
and per CCD pixel for a $\sim 15$~nm bandwidth. Our test bench is
similar to the one described in \cite{OConnorSPIE2016} for
LSST CCD acceptance test.

The CCD is operated within the vendor recommendations (see
table~\ref{table:CCDvoltages}): the drift field is created by applying
-70~V to the back substrate (namely the light entrance window of the
sensor). The CCD250 is a 4-phase device\footnote{The CCD 250 has 4
parallel phases and three phases in the serial register.}, and
during integration, we set 2 phases low and 2 phases high.
Even for short integrations, we do not run faster than $\sim$3~images
per min, which is the expected acquisition rate of LSST.  

\begin{table}[ht!]
  \begin{center}
    \begin{tabular}{llr}
      Voltage Name & Line & Voltage Value \\
      \noalign{\smallskip}
      \hline
      \noalign{\smallskip}
      \T Back Substrate bias     & BS        & -70.0~V\\
      \T Guard Diode             & GD        &  26.~V \\
      \T Output amplifier Drain  & OD        &  30.~V\\
      \T Output Gate             & OG        &   3.3~V\\
      \T Reset transistor Drain  & RD        &  18.~V\\
      \T Serial lines Low        & SL        &   0.6~V\\
      \T Serial lines High       & SU        &   9.8~V\\
      \T Parallel lines Low      & PL        &   0.06~V\\
      \T Parallel lines High     & PU        &   9.3~V\\
      \T Reset Gate Low          & RGL       &   -0.02~V\\
      \T Reset Gate High         & RGU       &  11.8~V\\
      \noalign{\smallskip}
      \hline
      \noalign{\smallskip}
    \end{tabular}
    \medskip
    \caption{Voltages used to operate the e2v-250 CCD.
      \label{table:CCDvoltages}}
  \end{center}
\end{table}

\begin{figure}[ht!]
\begin{center}
\includegraphics[width=1.05\linewidth]{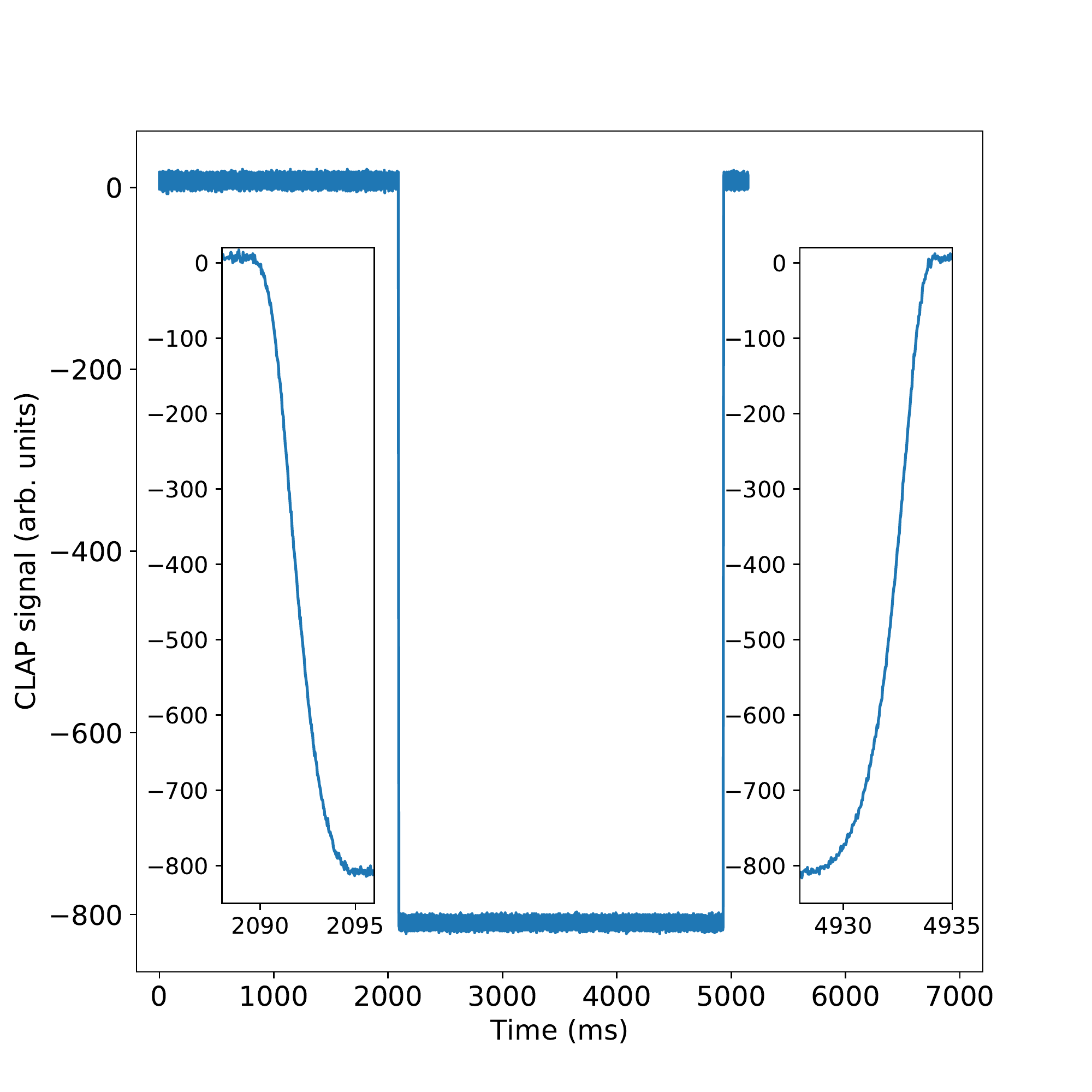}
\caption{Waveform acquired from our photodiode for a $\sim3\,\text{s}$ exposure.
The insets display the signal edges, which result from
the shutter motion rather than the bandwidth of the electronics.
\label{fig:clap}}
\end{center}
\end{figure}

\begin{figure}[htb!]
\begin{center}
\includegraphics[width=\linewidth]{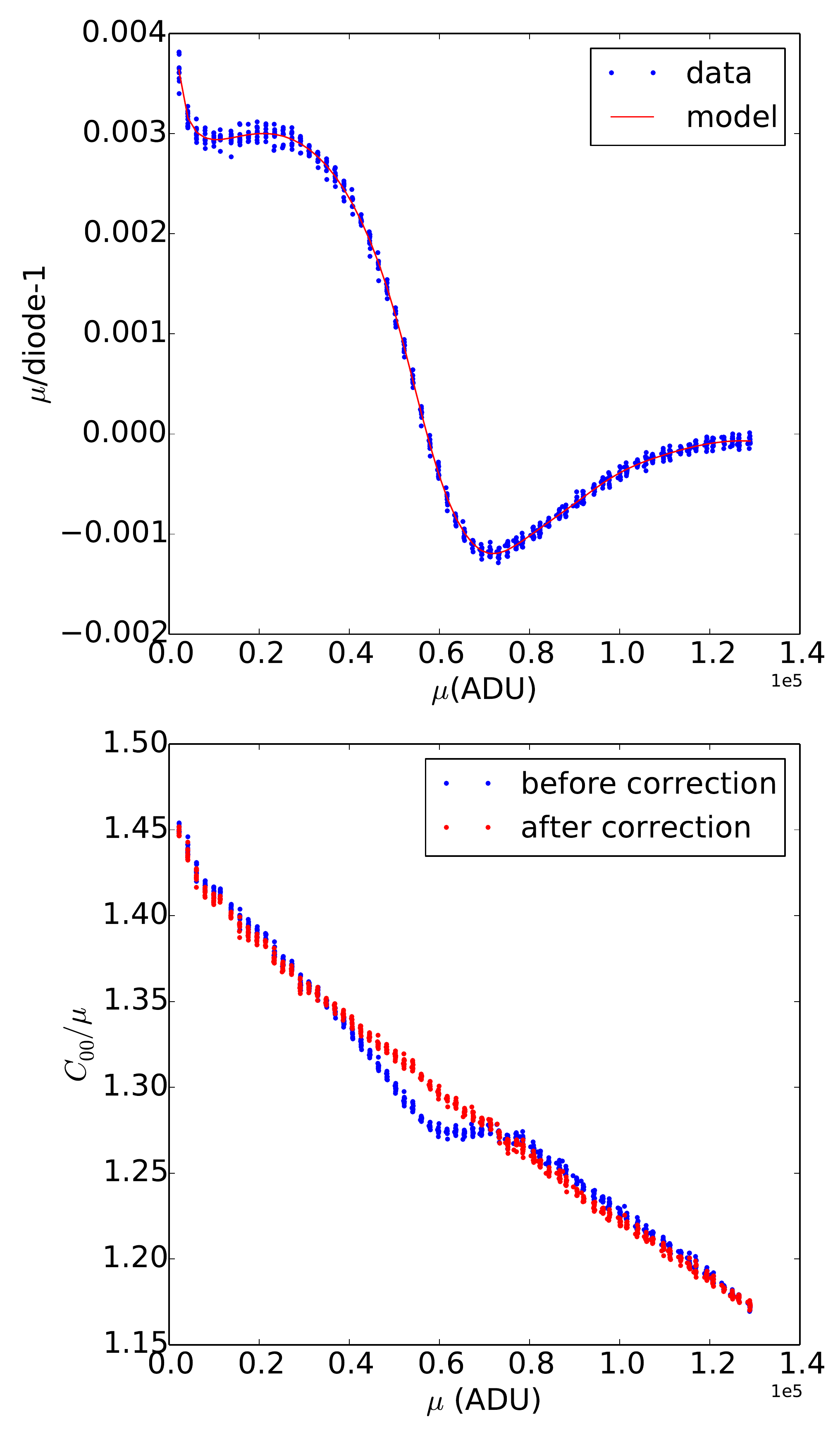}
\caption{Top: integral non-linearity of channel 8, and the spline
  model (with 14 knots) we use to correct for it. Bottom: $C_{00}/\mu$
  before and after non linearity correction. We see that the main
  distortion has disappeared.\label{fig:nonlin_plot}}
\end{center}
\end{figure}

\begin{figure}[hbt]
\begin{center}
\includegraphics[width=\linewidth]{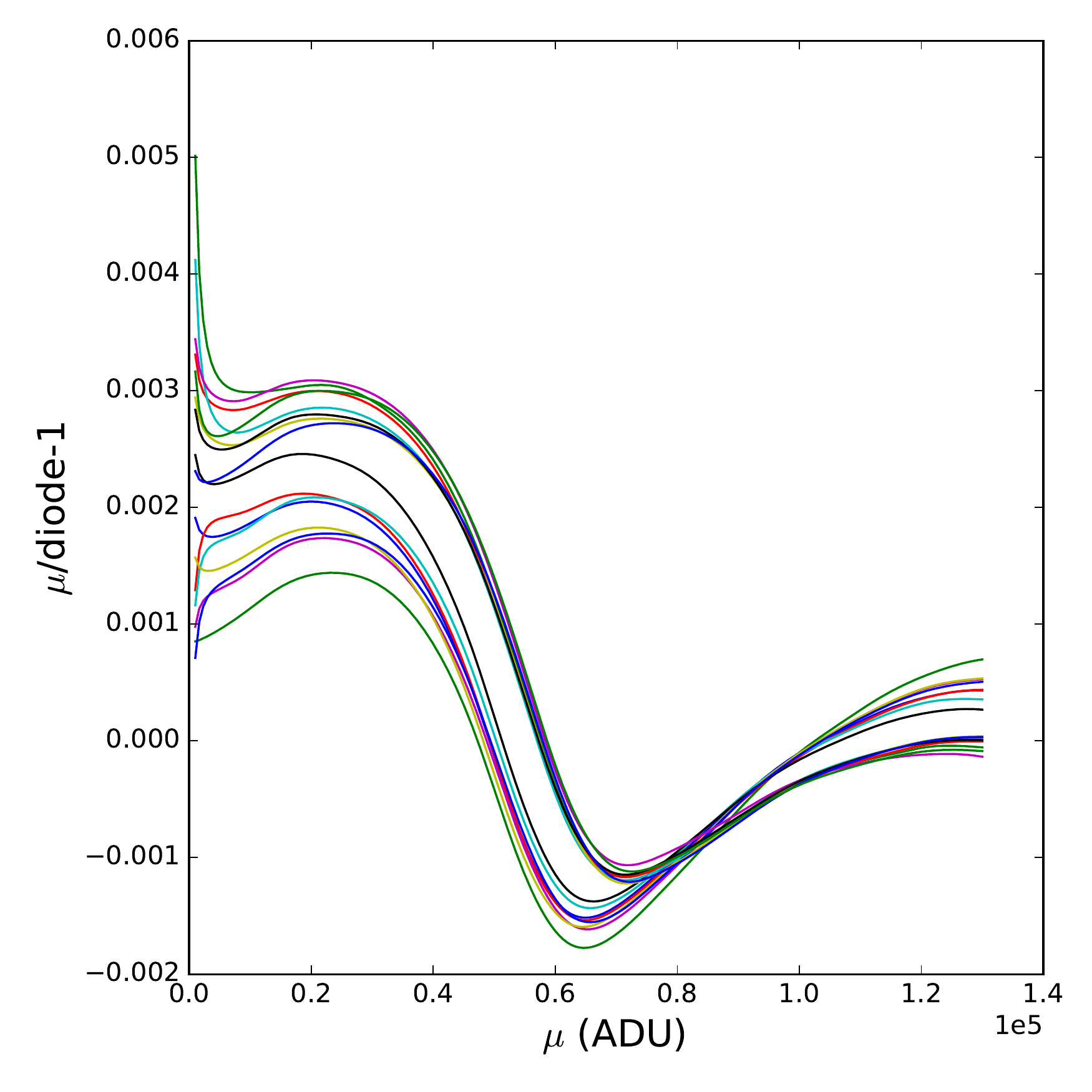}
\caption{Spline fits to integral non-linearity of the 16
    channels.  The main differences between various channels are the
    overall gains (the two groups correspond to the two different
    chips hosting the preamplifiers), and the low signal level behavior
    (presumably due to the on-CCD
    amplifier).\label{fig:nonlin_all_channels}}
  \end{center}
\end{figure}

In order to measure the non-linearity of the electronic chains, we use
the photodiode to measure the delivered light, integrating numerically the digitized
waveform, an example of which is shown in figure \ref{fig:clap}. Our video channels are affected by a small non-linearity, which we measure and correct (fig. \ref{fig:nonlin_plot}). In order to vary the integrated charge
  in the CCD image, we vary the open-shutter time at a somehow
  constant illumination intensity, ensured by our light source regulation.
  The photodiode hence delivers a current essentially independent of the exposure time,
  and our non-linearity measurement does not rely on its electronics being linear. We determine a non-linearity
correction for each channel and apply it to the input data\footnote{We first applied the correction to averages, variance and
  covariances, and it does not make any sizable difference}. We note that correcting
the variances for differential non-linearity is more important than
correcting the signal levels for integral non-linearity.
In the image series we consider
here, the pedestals do not vary by more than 10 ADU's, and so,
wondering if the non-linearity should be corrected before or after
pedestal subtraction is pointless. Figure \ref{fig:nonlin_all_channels}
displays the correction for the 16 channels, and one can notice that
the distortions are mostly similar. The prominent dip at roughly half the full range resembles in size and shape the non-linearity measured when testing the pre-amplifier, namely the ASPIC circuits. One may note that although our photodiode system allow us to measure the actual open-shutter time, we do not rely on this capability when establishing the non-linearity correction.
The plots in fig. \ref{fig:nonlin_plot} display the data of 10 successive acquisitions and allows one to visually judge the overall stability of the gain.

\subsection{Dataset and processing}
We use here 1000 pairs of flat field images (at 650 nm, as described above).
We refer here to pairs,
because all variance or covariance measurements are carried out on
pixel to pixel subtractions of pairs of flat fields at the same
intensity, in order to eliminate the contribution of illumination or
response non-uniformity to the measurements. The data is acquired 
as 10 successive sequences of flat pairs of increasing intensity, so that a slowly varying
gain (because e.g. of temperature variations) does not distort differently
low- and high-intensity images.

We first subtract the pedestal from the image, measured on the serial
overscan (ignoring the first 5 columns). We tried several approaches :
subtracting a global value per amplifier, and subtracting from each
line the median of its overscan.  The latter leaves residual
covariance of order of 2~\electron$^2$ along the serial direction, and we
eventually settled for a spline smoothing of the overscan along the
parallel direction. We then get residual covariances along the serial
direction of about $0.2$~\electron$^2$.

We clip 10 pixels on the four sides of the $512\times2002$ pixel area
of each channel, because at low and high x (i.e. serial
direction) and at low y (parallel direction),
the response of the sensor varies rapidly because of
distortions of the drift field near its edges. This clipping
is not required for all channels, but it allows us to safely compare
measurements from different channels.

The variances and covariances are computed from spatial averaging, and
using subtractions of image pairs of the same intensity eliminates the
contributions from non-uniformity of illumination or sensor sensitivity.
We however have to clip bad pixels, typically cosmetic defects
and ionizing particle depositions. To this end, we first clip
5~$\sigma$ outliers in each image of the pair, and then clip 4~$\sigma$
outliers on the subtraction. Toy Monte Carlo simulations show that
the bias on variance is $10^{-3}$ and twice as much for covariances.
We compensate this small bias in the analysis that follows.
With a cut at 3~$\sigma$ on the subtraction, the bias of the variance would be about 2.7 \%, 
and again twice as large for covariances.

One might wonder if the outlier clipping varies with signal level. We find on average that there are 8 more
  pixels (per channel of $10^6$ pixels) clipped at high flux ($10^5$~\electron) as compared with low flux. If we attribute these 8 pixels to the tails of a Gaussian distribution, the relative loss in variance is 1.4~$10^{-4}$. This corresponds to
  a loss of variance of about 10~\electron$^2$, while the effects we are discussing
  later are at least of the order of a few hundred. Furthermore, since
  the number or clipped pixels varies smoothly with signal level, most of the
  (small) induced bias is firstly absorbed by the gain.

We carry the computation
of covariances in the Fourier domain (the code is provided in appendix \ref{app:comp_fourier}), because it is faster than in direct space
as soon as one computes covariances for more than about $5\times 5$ separations,
and even less if fast Fourier transforms are computed for image sizes which
are powers of 2. 
In terms of symmetry, $C_{i,j} = C_{-i,-j}$, because these two expressions
are algebraically identical, as they involve exactly
the same pixel pairs. On the other hand, $C_{i,j}$ and $C_{i,-j}$
do not involve the same pixel pairs (if both $i$ and $j$ are non zero), but are expected, from 
parity symmetry, to be equal on average. So, we compute
both covariances (for $i\neq 0$ and $j \neq 0$) 
for sake of statistical efficiency, and report their average.
For small correlations, the statistical uncertainty of a covariance 
estimate is:
\begin{equation}
\sigma\left( \hat{C_{ij}} \right ) = V/\sqrt{N}
\end{equation}
for $i$ or $j$ non zero, where V is the variance of the pixel distribution, and N is the
number of pixel pairs used to evaluate the covariance. The uncertainty of the
variance estimation is twice as large. With $\sim$~$10^6$ pixels
per channel and image, and 1000 image pairs, we measure each
correlation ($C_{ij}/V$) with an (absolute) uncertainty of $\sim 3~
10^{-5}$, for each channel. The figure improves by a factor of 4 when
averaging over the 16 channels, and by $\sqrt{2}$ when both $i$ and $j$ are
non zero.

\subsection{Deferred charge}
\label{sec:deferred_charge}

Variance and covariances can be affected by contributions unrelated
the brighter-fatter effect, in particular imperfect charge transport:
if a small fraction of a charge belonging to a pixel is eventually
read out into its neighbor, a statistical correlation between
neighbors will build up. The quality of charge transport in CCDs is
commonly studied using ``overscans'', which correspond to clock and
read out cycles beyond the physical number of rows and/or columns.
Charges measured in these first overscan pixels are deferred signals, and
are commonly used to measure charge transfer inefficiencies (CTI), in
both the serial and parallel directions. 

\begin{figure}[hbt]
\begin{center}

\includegraphics[width=\linewidth]{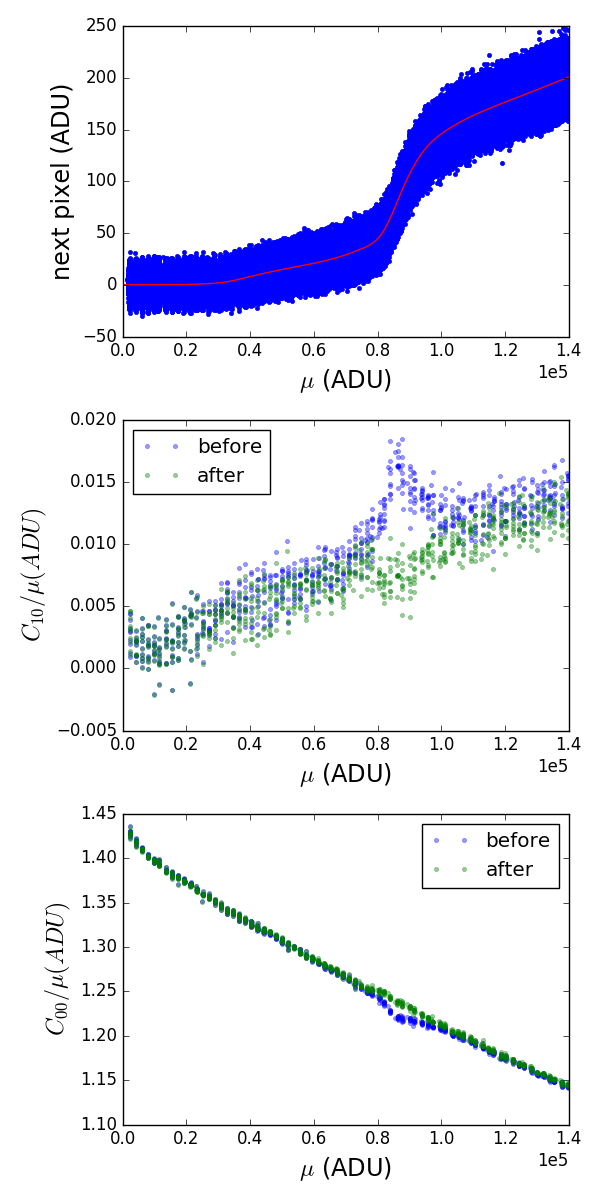}
\caption{The 3 plots refer to channel 10 only, after linearity correction.
  Top: content of the first serial overscan pixel as a function
  of flat level (data points and a spline fit). Middle: first serial
  correlation as a function of flat level, before and after placing
  the charge back, i.e., before and after deferred charge correction. Bottom: same as middle, but for the
  variance. In figure \ref{fig:nonlin_plot}, there is no obvious distortion of the variance curve after linearity correction because channel 8 is essentially free of deferred charge.  \label{fig:cti_plot}}
\end{center}
\end{figure}

\begin{table*}[hb!]
\begin{center}
\begin{tabular}{l|ccccc}
\TT         & $\chi^2_{full}/N_{dof}$ & $\chi^2_{2}/N_{dof}$ & gain & $a_{00}$ & RO noise \\ 
\hline
\TT value   & 1.23  & 4.04           & 0.713   & -2.377 $10^{-6}$         & 4.54     \\
\TT scatter & 0.10  & 0.27         & 0.020   & 0.032  $10^{-6}$         & 0.43   \\
\hline
\end{tabular}
\medskip
\caption{Statistics of PTC fits. The $\chi^2$ refer to the
  contribution of the variance terms, although the fit actually
  involves variances and covariances. The scatter is evaluated over the
  16 video channels. $\chi^2_{full}$ refers to the model of
  eq. \ref{eq:C_ij_mod}. The scatter of the gains essentially reflects
  different gains of the two integrated circuits hosting the
  pre-amplifiers. The observed spread of $a_{00}$ across amplifiers is
  several times larger than expected from shot
  noise. \label{tab:ptc_fit}}
\end{center}
\end{table*}

In figure \ref{fig:cti_plot}, we display the data of channel 10 only:
the top plot displays
the measured content of first serial overscan pixel
(after pedestal subtraction) as a function of the content
of the last physical pixel, while the two bottom plots display
the serial covariance, and the variance respectively. It is clear that
the rapid rise in deferred charge is associated to a peak in
the covariance, and a small trough in the variance.
To illustrate the relation between deferred charge and covariances, let us consider two
successive pixels, $p_0$ and $p_1$, each leaving some deferred charge
$\epsilon$ behind. $p_1$ becomes:
\begin{equation}
  p'_1 \equiv p_1 + \epsilon_0 - \epsilon_1
\end{equation}
We have:
\begin{align}
Cov(p'_1,p'_0) &= Cov(p_1,p'_0) + Cov(\epsilon_0,p'_0) \\
&\simeq Cov(p_1,p_0) + Var(p'_0) dE[\epsilon_0]/dp_0
\end{align}
So, the correlation between successive pixels follows the derivative of
the deferred charge.
Following a similar route, we get $Var(p'_1) \simeq Var(p_1) [1 - 2 dE[\epsilon_1]/d p_1]$, which means that we expect a variance deficit at the same
charge levels as the peaks in the serial correlation, 
that can be observed on the bottom plot of fig. \ref{fig:cti_plot}.
A toy simulation confirms that transferring a
signal-dependent charge to the next (time-wise) pixel produces peaks
on the $C_{10}$ vs $\mu$ curve at $\mu$ values where the deferred
charge varies rapidly. In the case where a constant fraction is transferred
to the next pixel (possibly negative, at some point of the video chain), this causes a
constant correlation offset, so mostly a linear contribution to the covariance
which does not exist in the covariance models discussed above.

We do not have a clear physical model of what causes the deferred
serial charge to increase rapidly at specific signal values. This
non-linearity seems to exclude trailing signals in the electronic
chain. Not all channels have clear $C_{10}$ peaks (for example channel
8 displayed in figure \ref{fig:nonlin_plot} is essentially free of
such effects) and different channels have peaks at different $\mu$
values. These peaks remain located at the same positions expressed in
number of electrons when using an other readout board or when altering
the gain at the CCD level. This indicates that their cause lies in the
sensor itself. The sizes of these peaks are similar in all sub-regions
  of the image section corresponding to a given channel. We hence attribute those to some effect happening in 
  the pre-scan pixels of the serial register, that all collected
  charges have to traverse.  We finally observe that the height of
these peaks varies with the {\it parallel} gate voltages, which may
influence the electric field in the vicinity of the serial
registers. We are currently studying a possible mitigation of these
peaks by altering the parallel clock levels.

We have modeled the average deferred serial charge by fitting a spline curve
to the measurements (fig. \ref{fig:cti_plot}, top), for each channel
separately, based on the content of the last physical pixels  and the first overscan pixels. Using this simple modeling, we pre-process the images by
placing the (average)
deferred charge back into the previous pixel. Once the correction is
done, the peaks in the $C_{10}$ curve vanish as well as their
counterparts on the variance curve (fig. \ref{fig:cti_plot},
middle and bottom). We do not know why this simple correction procedure slightly
over-corrects the $C_{10}$ peak (fig. \ref{fig:cti_plot} middle).
In this figure, one can notice that the correction of deferred charge
reduces the slope of $C_{10}/\mu$ by roughly 10\%, and inspecting the model
of equation \ref{eq:C_ij} shows that this slope drives
the $a_{10}$ coefficient value. 

We note that the correction procedure we used assumes that all columns
are affected in the same way, i.e. that all charges go through the
defect causing deferred charge. This is justified by covariances
measured separately for different set of columns being similar, and by
the ``pocket-pumping'' technique (see e.g. \S 5.3.4 in
\citealt{JanesicBook01}) not unveiling traps in the serial register.

The parallel transport of this sensor is far better than the serial
one: we typically measure a few ADUs of deferred signal close to
saturation, as compared to more than 100~ADUs in the serial
direction. We hence decided to not correct for this small effect. We
convert the measured values of deferred charge into the commonly used
CTI figure: the channel displayed in fig. \ref{fig:cti_plot} has
a high signal serial CTI of $200/1.4~10^5/522 \simeq 3~10^{-6}$, where
$522$ is the number of pixels in the serial register. The highest accepted
serial CTI for LSST is set to $5~10^{-6}$. The parallel CTI of this sensor
is better than $2~10^{-8}$at high flux.

We finish this section by one more puzzling observation: the channel
displayed in fig. \ref{fig:cti_plot} has a gain of approximately 0.7
(el/ADU), meaning that the overscan pixel reaches about 130 electron
at the maximum $\mu \simeq 10^5$~\electron. At this highest value, we measure
a variance of $\sim$~60~\electron$^2$, of which typically 25 are expected
from read noise (as measured at the other end of the curve). So, at
$10^5$ \electron, the variance of the deferred charge ($\sim
35$~\electron$^2$) is less that one third of its average ($\sim 130$),
i.e. much less than expected from Poisson statistics.  At all values,
the measured deferred charge fluctuates much less than expected from
Poisson fluctuations and read noise. A physical model for this
deferred charge would have to accommodate this observation, but
this goes beyond the scope of this paper.

\section{Analysis of the Photon Transfer Curve and  covariance curves}
\label{sec:analysis}

\begin{figure}[htb]
\begin{center}
\includegraphics[width=\linewidth]{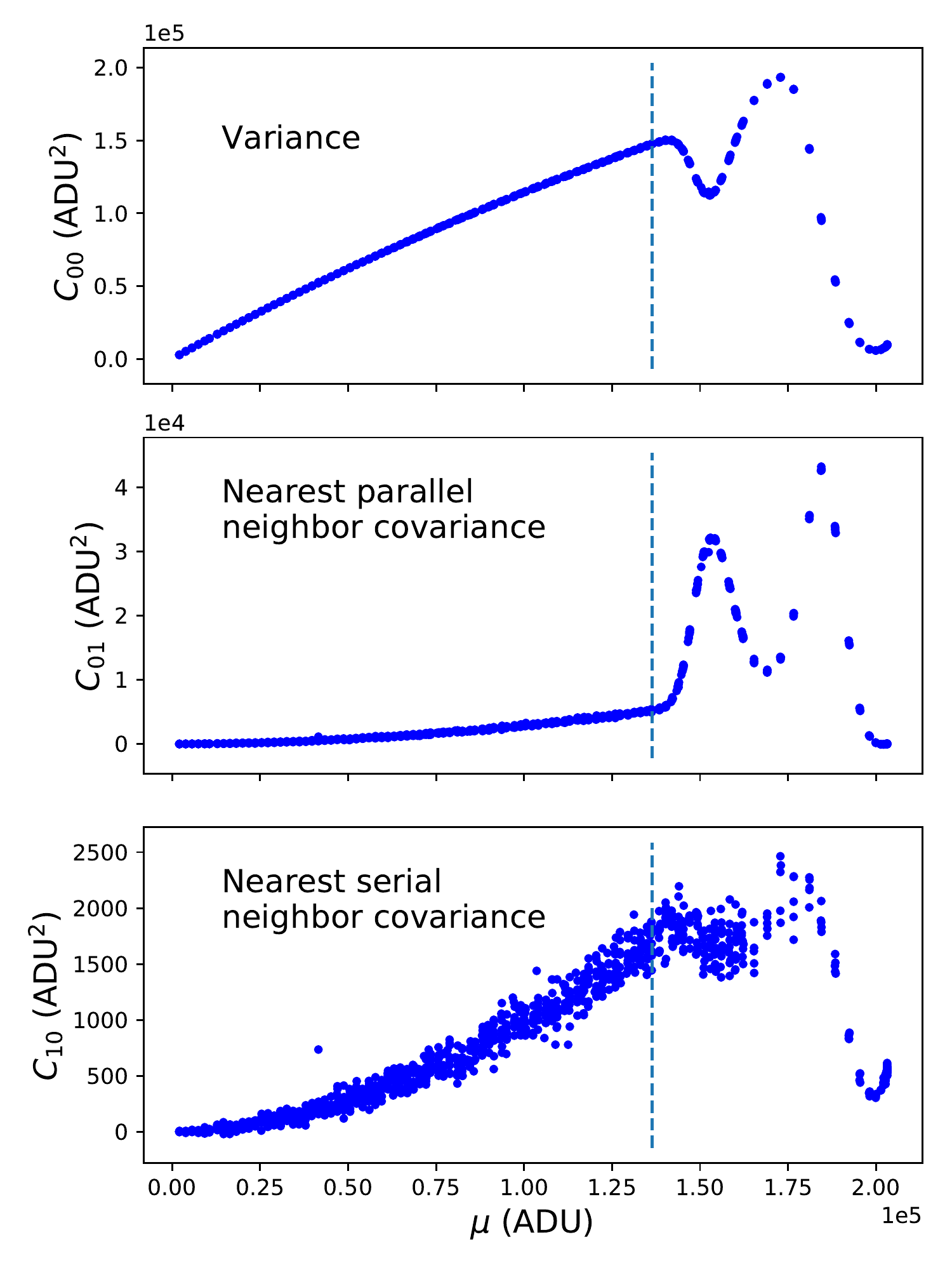}
\caption{Variance and two nearest covariances as a function of flat field average,
  in order to illustrate the saturation level (in channel 0). The three curves change
  behavior around 1.5~$10^5$ ADUs. The drawn fitting limit is about 10\% below this value, and corresponds to $10^5$ \electron, at a gain of 0.733.
\label{fig:satur_plot}}
\end{center}
\end{figure}

\begin{figure}[htb]
\begin{center}
\includegraphics[width=\linewidth]{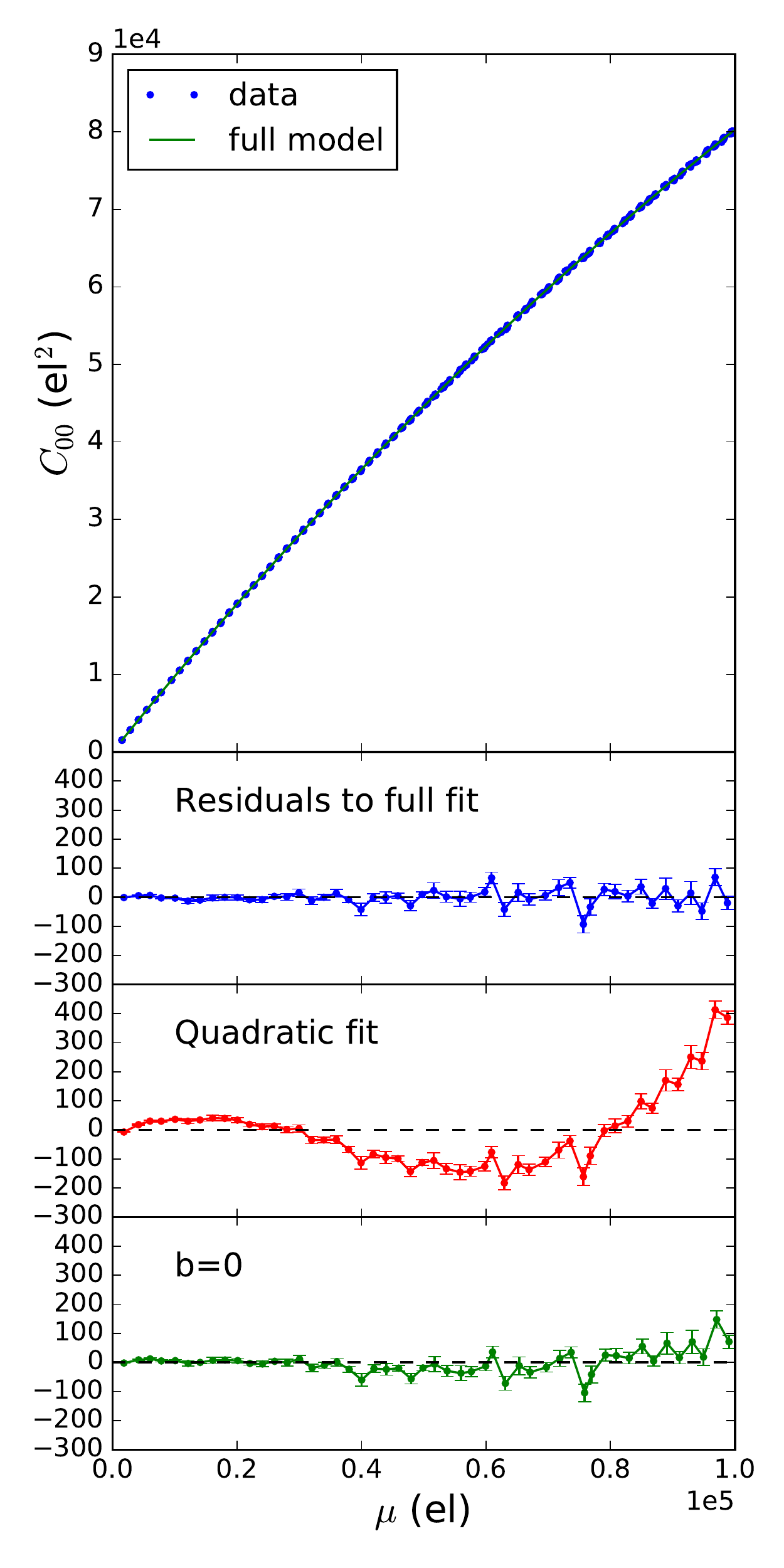}
\caption{PTC of channel 0 fitted with three different models : the full model, a quadratic fit, and the ``simple model'' with $b=0$. The
  bottom plots display the fit residuals. The full model
  and the $b=0$ model describe well the data, although the $\chi^2$ is 10\%
  higher for the latter. We believe that part of the fit residuals
  are artifacts caused by the non-linearity and deferred charge corrections. \label{fig:ptc_fit_plot}}
\end{center}
\end{figure}

We now confront the models presented above to our data set. We fit the
$C_{ij}$ vs $\mu$ relation with two models, eq. \ref{eq:C_ij} and
\ref{eq:C_ij_mod}, up to terms in   $(\mu g)^3$  in the bracket.
The fitted parameters are the gain, the $a_{ij}$ and
$n_{ij}$ quantities, and when applicable $b_{ij}$.
For $\mu$, we simply use the average between the two members of the image pair.

We fit 64 $C_{ij}$
curves with $i<8$ and $j<8$, because beyond this separation, the
  signal is much smaller than shot noise, for our data sample.
We fit separately each readout channel,
using \verb+scipy.optimize.leastsq+, without binning the data. The uncertainties
are derived from shot noise. 
We iteratively reject outliers measurements at the 5 $\sigma$ level, where
the cut is derived from the scatter of the residuals. We initialize the parameters from separate polynomial fits to each separation. The fit takes a few seconds (on a core i7 CPU) for
about 50,000 measurements. 
In the figures we display here, we
generally choose to display $C_{ij}/\mu$ rather than correlations
($C_{ij}/C_{00}$), because the latter involve the measured variance,
itself affected by the brighter-fatter effect.

We first have to decide up to which charge value we should fit.
Variance and nearest neighbor covariances change behavior at similar
filling levels, as shown on fig. \ref{fig:satur_plot}, and it is then
fairly easy to define a saturation. 
We arbitrarily choose a margin of about 10\% below this saturation, which
makes $10^5$ \electron. All fits that are described here
are carried out to up to this level. Applying this cut involves some iterative
procedure since it requires to know the gain, which requires some crude
fit of the PTC (we use a third degree polynomial).

\begin{figure}[htb]
\begin{center}
\includegraphics[width=\linewidth]{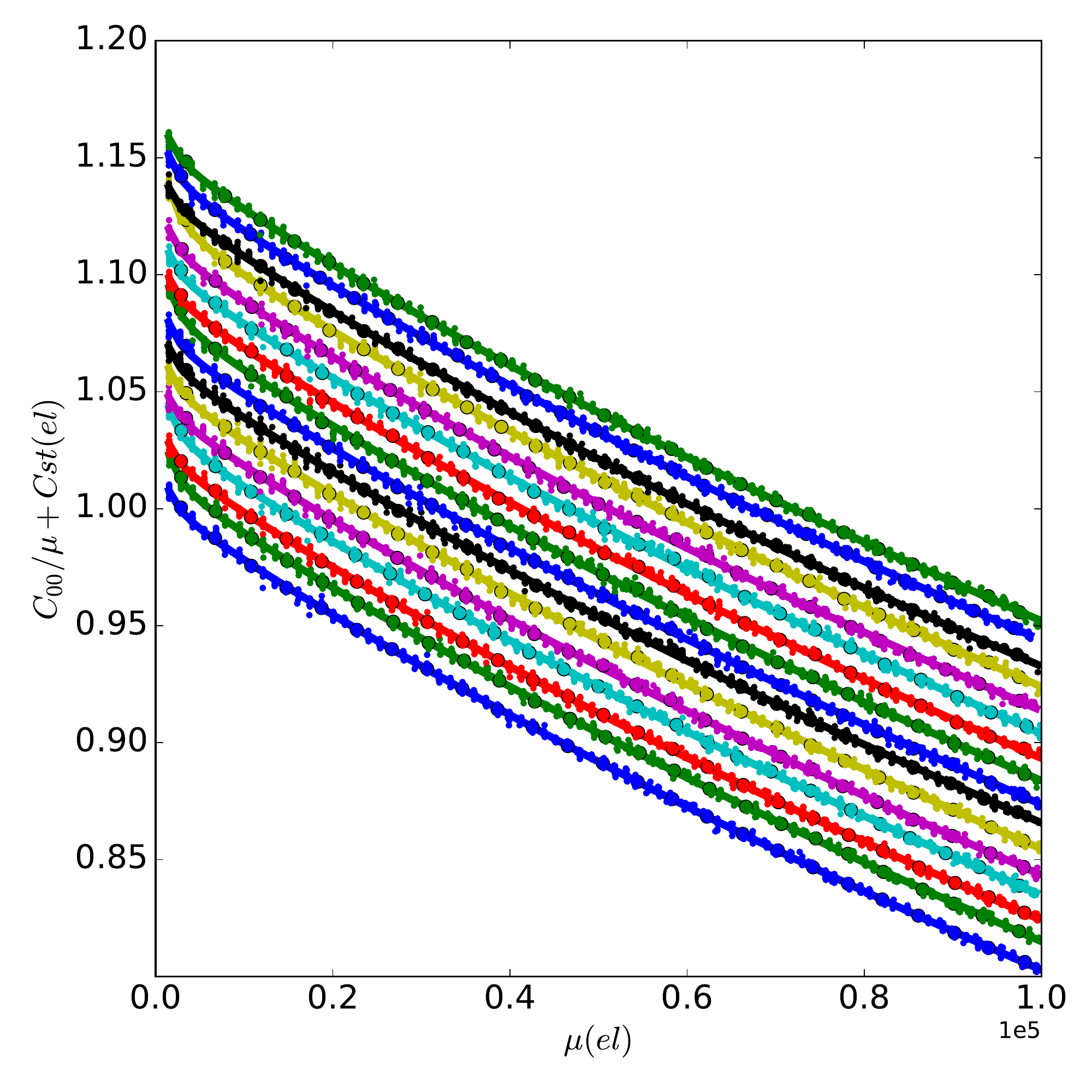}
\caption{PTC of all channels (namely $C_{00}/\mu$), offset by 0.01 times
  the channel number. The curves going through the data points correspond to the full fit. The statistics associated to these fits are provided in table \ref{tab:ptc_fit}. \label{fig:C00_plot}}
\end{center}
\end{figure}

Figure \ref{fig:ptc_fit_plot} displays the PTC fit result
of channel 0, and residuals to the two models above, and to a parabolic
fit. The inadequacy of the parabolic fit is clear, but allowing for a
non-zero $b$ term does not dramatically improve the fit quality at least visually, although the $\chi^2$ goes down by $65$ on average (over channels)
for a single extra parameter. The PTCs of the 16 channels are shown in Figure \ref{fig:C00_plot} and Table \ref{tab:ptc_fit} details the outcome
of the data fits regarding the PTC itself (i.e. $C_{00}$). We find that
the model from eq. \ref{eq:C_ij_mod} describes the data fairly well, although $\chi^2$
are higher than expected from shot noise, which we attribute to imperfections
of the non-linearity and deferred charge corrections, which are both
localized in $\mu$ because we use highly flexible functions to model both.
The excess in $\chi^2$ corresponds to offsets at the $2\ 10^{-4}$ level r.m.s.

\begin{figure}[h!]
\begin{center}
\includegraphics[width=\linewidth]{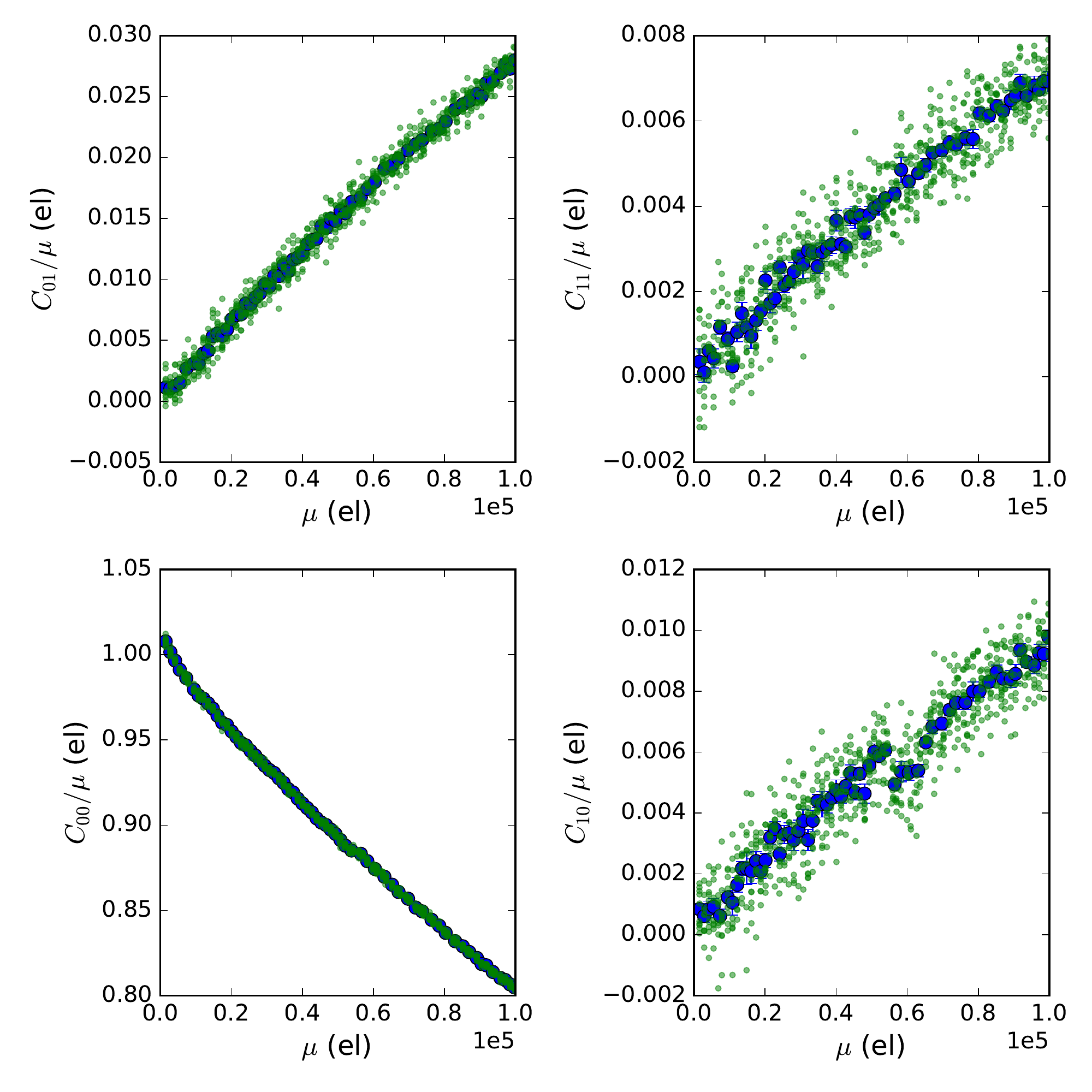}
\caption{Nearest neighbor covariances for channel 10, and
  variance (divided by $\mu$). We display both the individual measurements and
  binned values. All quantities are expressed in electron units. The spread is similar for all plots but appears differently because scales are different. The
  wiggle on $C_{10}$ around $\mu=0.6\ 10^5$ is a remainder of the deferred charge
  correction (\S \ref{sec:deferred_charge}). It is clear that the data does not follow straight lines.
\label{fig:cov_exposure_plot}}
\end{center}
\end{figure}

\begin{figure}[h!]
\begin{center}
\includegraphics[width=\linewidth]{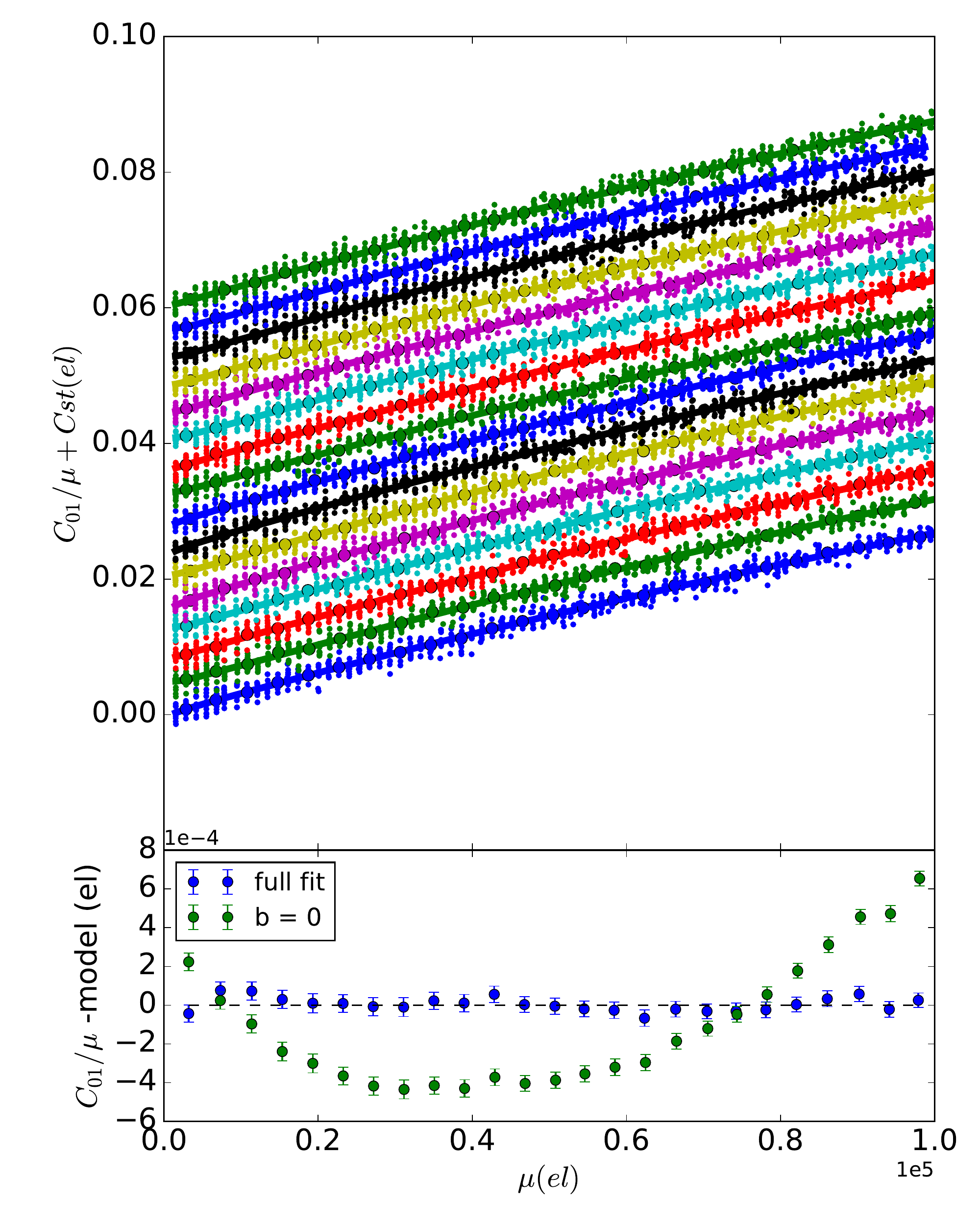}
\caption{Fits of $C_{01}/\mu$ as a function of $\mu$ for the 16 channels.
  On the top plot, the data and model (the full model of eq \ref{eq:C_ij_mod})
  have been offset by $0.004\times c$ where $c$ is the channel number. We plot, both individual measurements
and averages at the same intensity, where the error bars reflect
the scatter (compatible with the expected shot noise).
The bottom plot reports the binned average residuals to both fits.
\label{fig:C01_fit_plot}}
\end{center}
\end{figure}

The next step is to study the covariance fits. We first illustrate the
data by displaying the variance and 3 nearest neighbor covariances in
fig. \ref{fig:cov_exposure_plot} for channel 10, which is affected by
a varying deferred charge.  All covariance measurements have similar
statistical uncertainties, so the best relative uncertainties are
obtained for the highest covariances, i.e. for the nearest neighbors.
We first display the fit of the largest covariance i.e. $C_{01}$
(which is not affected by deferred charge). We display the fits
results in fig. \ref{fig:C01_fit_plot} to each channel separately and
report the statistics of these fits in table
\ref{tab:C01_curve_stats}.  Visually, it is very clear that $b \neq 0$
is required. Allowing for $b \neq 0$, the $\chi^2$ goes down by 106 on average
over channels, for a single extra degree of freedom. Taken at face value,
the sense of $b_{01}$ is that the effect of a given charge on its nearest
parallel neighbor pixel is 17
\% larger at $\mu=10^5$ \electron\  than linearly extrapolated from low-average
data. The spread of this 17 \% figure, as measured over the 16 channels,
is only 3\%.
\begin{table}[h]
\begin{center}
\begin{tabular}{l|ccc}
\TT  & $a_{01}$ &  $b_{01}$ &  $\chi^2/N_{dof}$ \\
  \hline
\TT  value &  3.32 $10^{-7}$ &  1.71 $10^{-6}$ &  1.03\\
\TT  scatter & 0.06 $10^{-7}$ & 0.29 $10^{-6}$ &  0.04 \\
  \hline
\end{tabular}
\medskip
\caption{Statistics of $C_{01}$ curve fits (mean and r.m.s. scatter over read out channels). The scatter of $a_{01}$ over channels is about twice as large as expected from shot noise. \label{tab:C01_curve_stats}}
\end{center}
\end{table}

\begin{figure}[h!]
\begin{center}
\includegraphics[width=\linewidth]{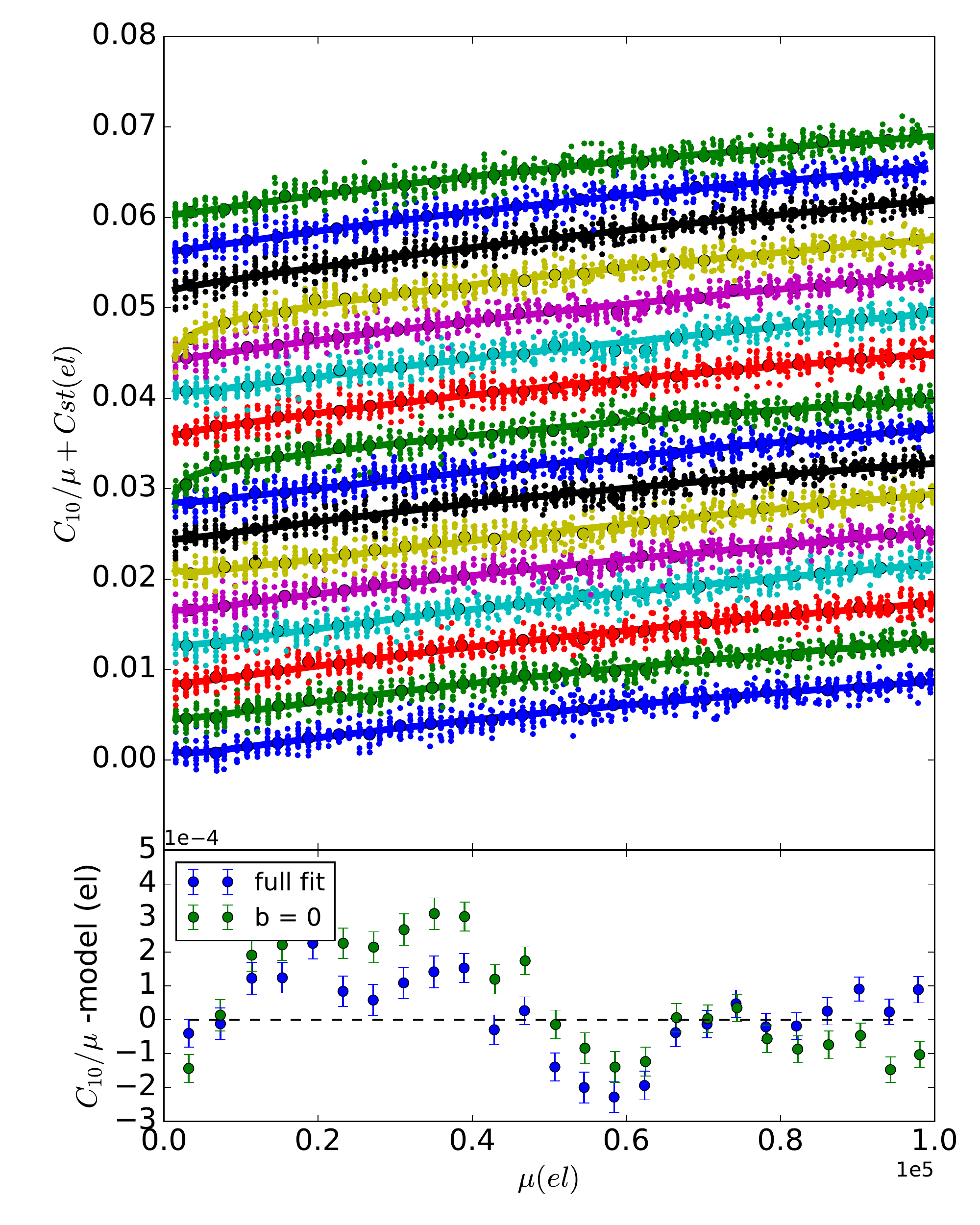}
\caption{Fits of $C_{10}/\mu$ as a function of $\mu$ for the 16 channels.
  On the top plot, the data and model (the full model of eq \ref{eq:C_ij_mod})
  have been offset by $0.004\times c$ where $c$ is the channel number. We plot, both individual measurements,
and averages at the same intensity, where the error bars reflect
the scatter (compatible with the expected shot noise). The down turn at low
flux of channels 8 and 12 is accommodated in the fit by the $n_{10}$ term in the model equation \ref{eq:C_ij_mod}.
The bottom plot reports the binned average residuals to both fits.
\label{fig:C10_fit_plot}}
\end{center}
\end{figure}

The nearest serial neighbor covariance $C_{10}$ is much noisier
than $C_{01}$, because the covariance is about 2.5 times smaller, and
wiggles have survived the deferred charge correction. We however
display the fit results in figure \ref{fig:C10_fit_plot} and the fit
statistics table in table \ref{tab:C10_curve_stats}. The scatter of the
measured coefficients is again larger than expected from shot noise, by about
a factor of 3 for $a_{10}$, although the $\chi^2$ of the fits are close to
the statistical expectation. The $\chi^2$ improvement due to $b \neq 0$
is 17 on average over channels. Note that $b_{10}$ is negative, i.e., the
anisotropy of nearest neighbor covariances seems to increase with
charge levels.

\begin{table}[h]
\begin{center}
  \begin{tabular}{l|ccc}
\TT    & $a_{10}$ &   $b_{10}$ &  $\chi^2/N_{dof}$ \\
\hline
\TT value &  1.26 $10^{-7}$ & -1.77 $10^{-6}$ &  1.03\\
\TT scatter & 0.08 $10^{-7}$ &  0.97 $10^{-6}$ &  0.07\\
\hline
\end{tabular}
\medskip
  \caption{Statistics of $C_{10}$ curve fits (mean and r.m.s. scatter over read out channels). The scatter of $a_{10}$ over channels is about three times larger than expected from shot noise. \label{tab:C10_curve_stats}}
\end{center}
\end{table}

\begin{figure}[h!]
\begin{center}
\includegraphics[width=\linewidth]{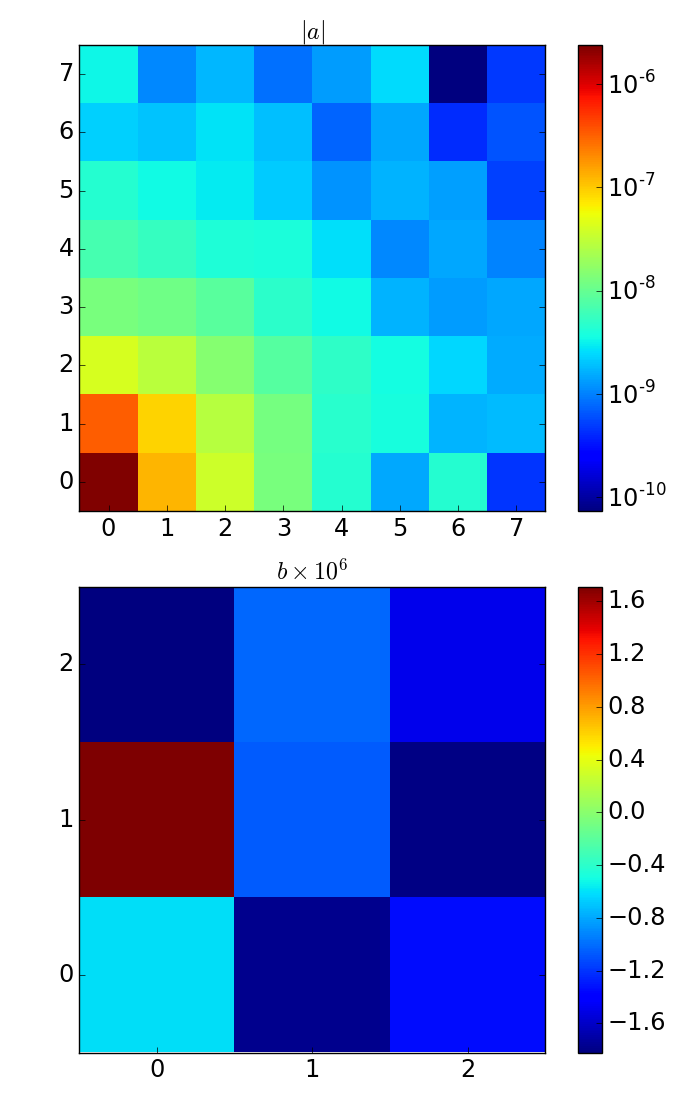}
\caption{Color display of $a$ and $b$ arrays fits, averaged over channels.
The parallel direction is vertical.\label{fig:a_and_b}}
\end{center}
\end{figure}

\begin{figure}[h!]
\begin{center}
\includegraphics[width=\linewidth]{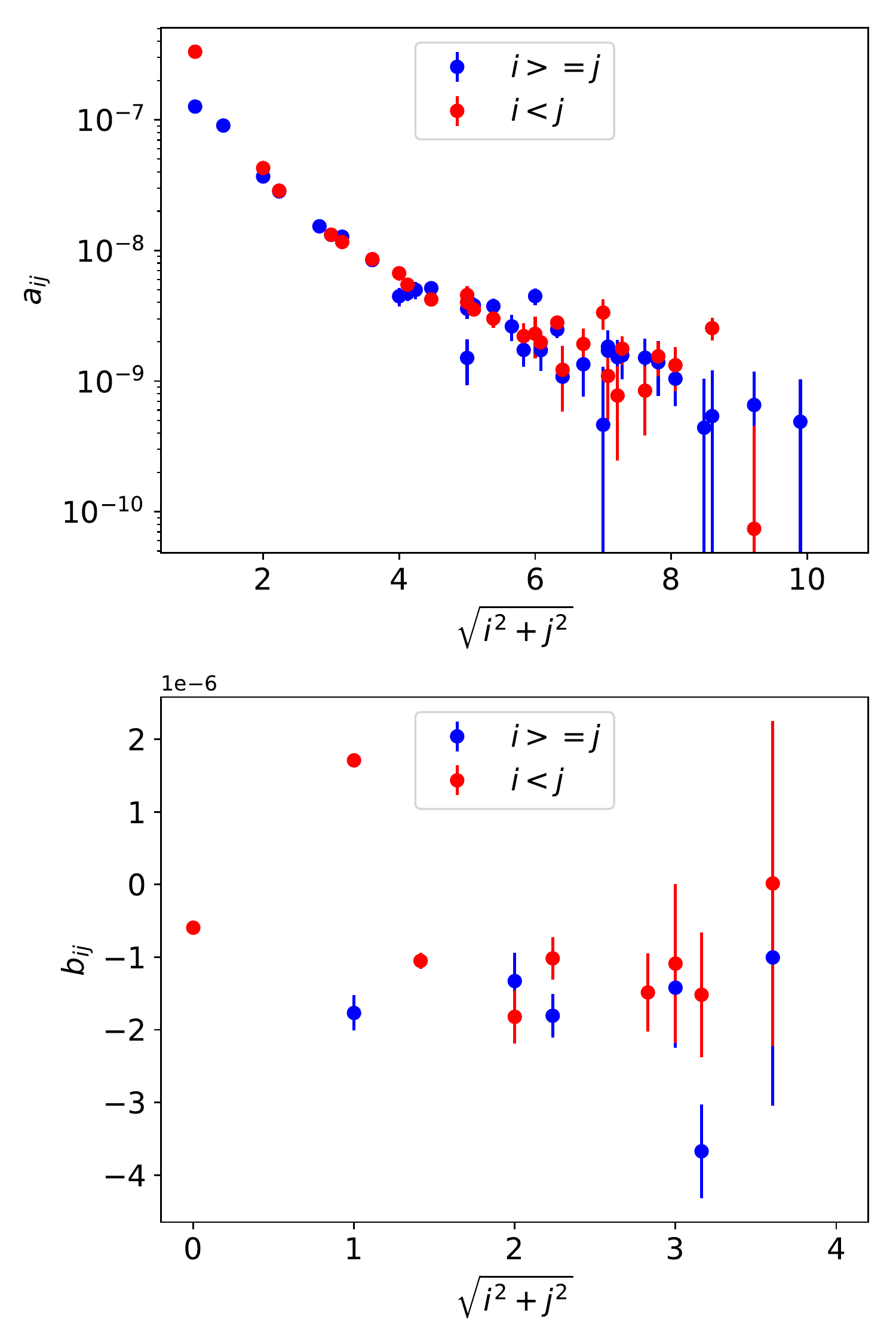}
\caption{Values of $a$ and $b$ arrays fits, averaged over channels,
  as a function of distance. Error bars represent the uncertainty of the
  average, derived from the observed scatter. One can notice that
  $b$ values are mostly negative.\label{fig:ab_vs_dist}}
\end{center}
\end{figure}

Figure \ref{fig:a_and_b} displays the fitted $a$ and $b$ values,
averaged over channels. At increasing distances, $a$ decays rapidly
and becomes isotropic, as visible in fig \ref{fig:ab_vs_dist} (top).
In figure \ref{fig:ab_vs_dist}, we display the $a$ and $b$ values
averaged over channels, as a function of distance, with error bars
representing the uncertainty on the average, derived from the observed
scatter. One can deduce that, with our data, individual $a$ values are
measured down to distances of about 8 pixels, and $b$ values down to
about 3 to 4 pixels. For a fit to some electrostatic model, one could
use even farther measurements with proper weights. As already noted
$b_{01}$ and $b_{10}$ are found of opposite signs, and examining the
bottom plot of fig. \ref{fig:ab_vs_dist}, one observes that $b$ values
are mostly negative, with the notable exception of $b_{01}$. A
positive $b_{01}$ can be due to the cloud getting broader in the
parallel direction as charge accumulates. We can then imagine two
distance-independent contributions to $b$: first the decay of the
drift field due to stored charges (sometimes referred to as space
charge effect), or the electron clouds center of gravity gradually
changing distance to the parallel clock stripes as charge accumulates.
The first effect would cause positive $b$ values, and hence, if
present, has to be sub-dominant. If one attributes the negative $b$
to the second effect, the wells have to move {\it closer} to the
clock stripes as charge accumulates.  Regarding the stored charge
diminishing the drift field (or space charge effect), one should note
that even if this effect contributes to flat field electrostatics, it
is in practice mostly absent from science images which usually have
low to moderate sky background levels.

\begin{figure}[h!]
\begin{center}
\includegraphics[width=\linewidth]{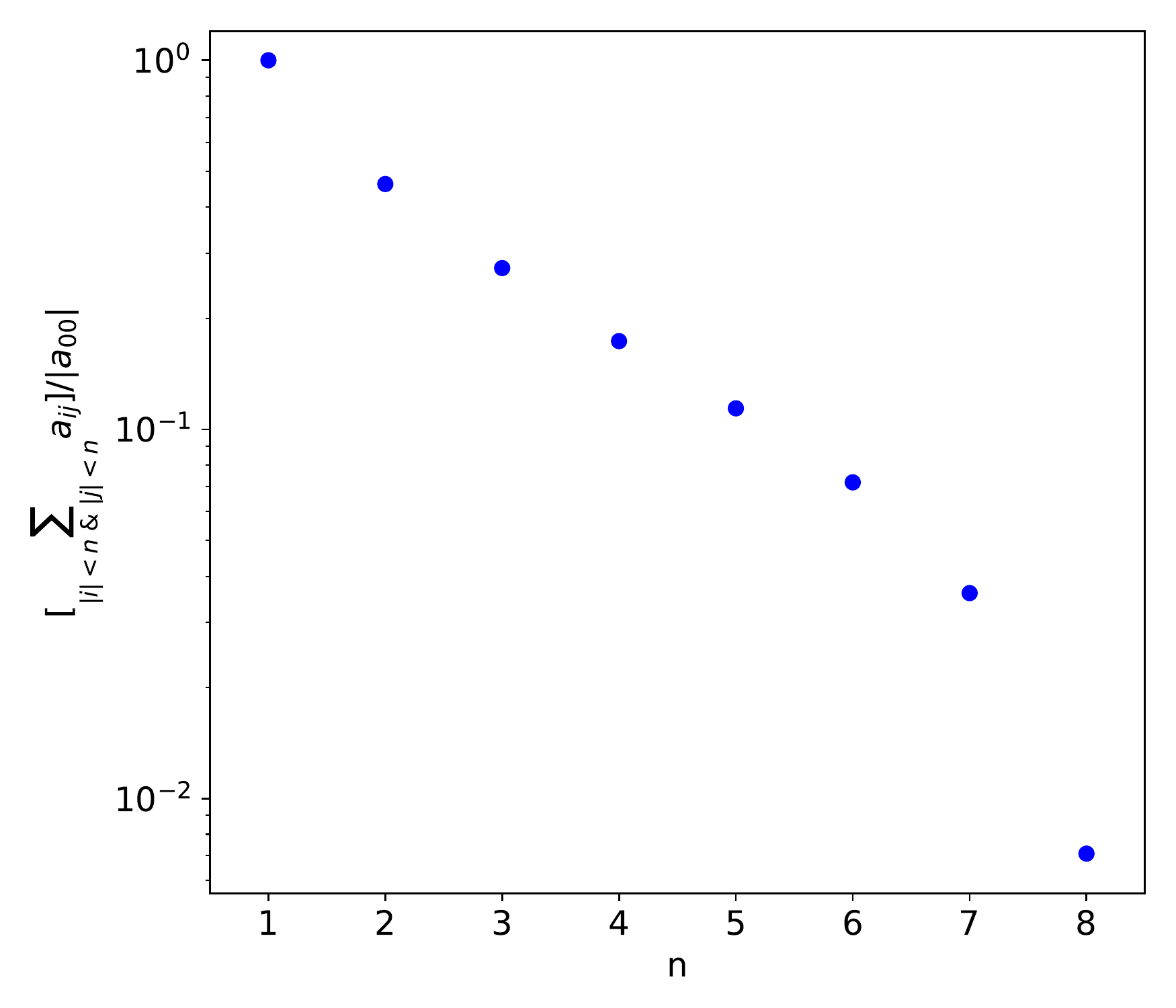}
\caption{Cumulative sum of $a_{ij}$ as a function of maximum separation. This plot displays the average over channels.\label{fig:plot_a_sum}}
\end{center}
\end{figure}

One final analysis point is the empirical verification of the sum rule
in eq. \ref{eq:sum_rule}, that we have not enforced in the fit. On average
over channels, we find:
\begin{equation}
  \sum_{-8<k,l<8} a_{kl}   =  -1.6 \ 10^{-8} \pm 5.7 \ 10^{-8}
\end{equation}
where the uncertainty reflect the scatter (to be divided by 4 for the
uncertainty of the average). To set the scale, $a_{00} \simeq -2.4\ 
10^{-6}$, i.e. the sum rule is satisfied down to the \% level, thus
indicating that the effects we measure are mostly due to charge
redistribution. Figure \ref{fig:plot_a_sum} displays the sum as a function
  of maximum separation, and shows that, for our sensor, about 1\% of the area lost (or gained) by a pixel due to its charge content is
  transferred to pixels located 7 or more pixels away.

We finish this paragraph by a short technical note about the fitting
approach: we have considered fitting in Fourier space rather than in
direct space because the expression of the model does not involve
convolutions. This is however the only benefit over direct space:
first, starting from covariances in direct space, one has to Fourier
transform those, and the mandatory zero padding correlates the
transformed data (which would otherwise be statistically independent
because measured covariances have uncertainties independent of
the separation). Second, one can think that different ``reciprocal separations'' in Fourier space
could be fitted independently, but the common gain parameter still imposes to
fit all the data at once. Finally, it is trivial to concentrate on
small distances in direct space, but selecting large-k modes in the
power spectrum of image pairs requires manipulating a lot of data.
So, we have not been able to devise a simpler framework in Fourier
space.

\subsection{Inaccuracies of simpler analyses of covariance curves}
\begin{figure}[h!]
\begin{center}
\includegraphics[width=\linewidth]{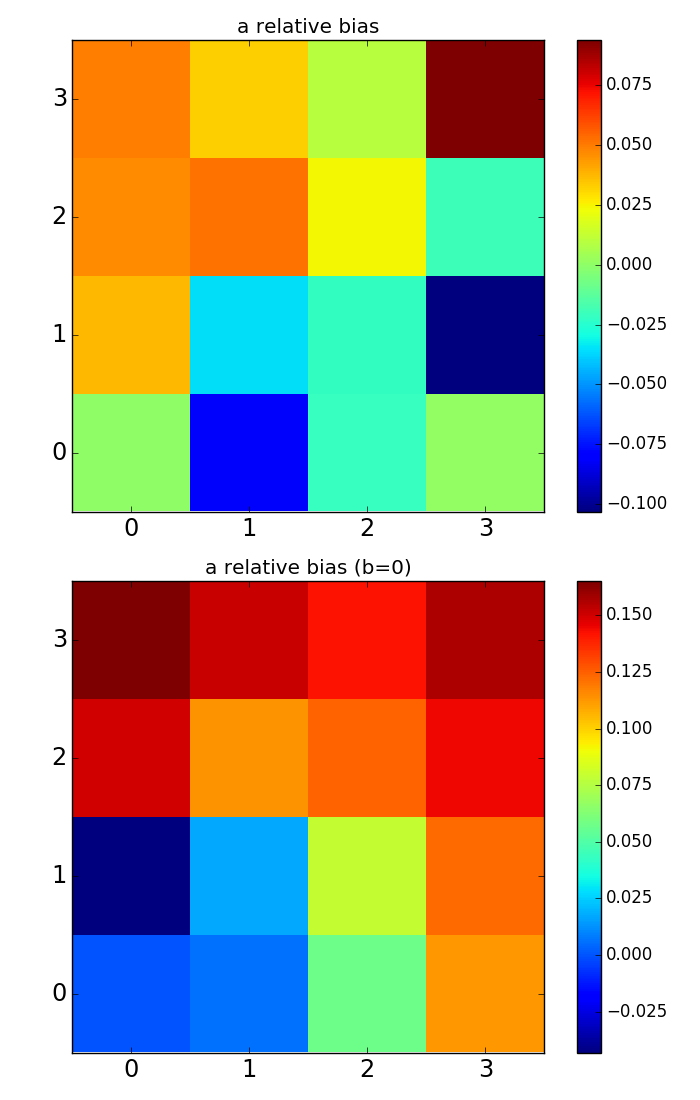}
\caption{Relative systematic bias of the commonly used method that estimates
  $a_{ij}$ from the slopes of correlations, using a single
  illumination value at 75~ke. The top figure is for the full model,
  and the bottom one is for $b=0$. In the top case the nearest neighbors
  $a_{ij}$ are biased by +8\% and -4\% respectively. For the bottom case,
  there is a global overestimation of the effect, increasing with distance.
  \label{fig:a_bias}
}
\end{center}
\end{figure}

Many works in the literature using pixel covariances to constrain (and correct)
the brighter-fatter effect \citep{Antilogus14, Guyonnet15, Gruen-PACCD-15, Coulton-17} assume that pixel correlations are linear with respect to
illumination level, i.e. :
\begin{equation}
  C_{ij} = a_{ij} V \mu \label{eq:simple-cov}
\end{equation}
where $V$ generally represents the measured variance $C_{00}(\mu)$, and sometimes only one intensity (preferably high, in order to optimize the S/N) is being
measured. Using the expression above is a simplification, as compared
to equations \ref{eq:C_ij} and \ref{eq:C_ij_mod}. We evaluate the biases caused
by this simplification alone, using the model fits reported in the
previous section. To this end, we compare
$a_{ij}$ as extracted from the simplified equation above with the one of
the model used to evaluate $C_{ij} $ and $C_{00}$.
As in the previous section, we study both the full fit result and the
$b=0$ case. 
In figure \ref{fig:a_bias}, we display the relative difference between
$a_{ij}$ evaluated using the above equation and the model value, for a
measurement that would be performed at 75~ke
(a value representative of the data studied in \citealt{Guyonnet15}),
averaging over the 16 video channels.
We see that for the full fit ($b \neq 0$), the nearest neighbor coefficients
are offset by -8\% and +4\% respectively. Using these biased inputs to
correct science images not only compromises the
measurement of the size of the PSF, but more importantly causes
spurious PSF anisotropy. For the case
where $b=0$, we see that essentially all $a_{ij}$ are overestimated
(except $a_{01}$, biased by -4\%), with a relative bias increasing
with separation, up to more than 10 \% at distances of $\sim 3$. For both models,
the bias arises from neglecting (or approximating) terms beyond $\mu^2$
in equations \ref{eq:C_ij} and \ref{eq:C_ij_mod}.
Depending on the considered sensor, it is plausible that the 
limitations of the brighter-fatter effect corrections observed empirically
(e.g. \citealt{Guyonnet15, Mandelbaum-HSC}) are partly due to
the oversimplification implied by eq. \ref{eq:simple-cov}.

\subsection{Non-electrostatic contributions to covariances}
\begin{figure}[h!]
\begin{center}
\includegraphics[width=\linewidth]{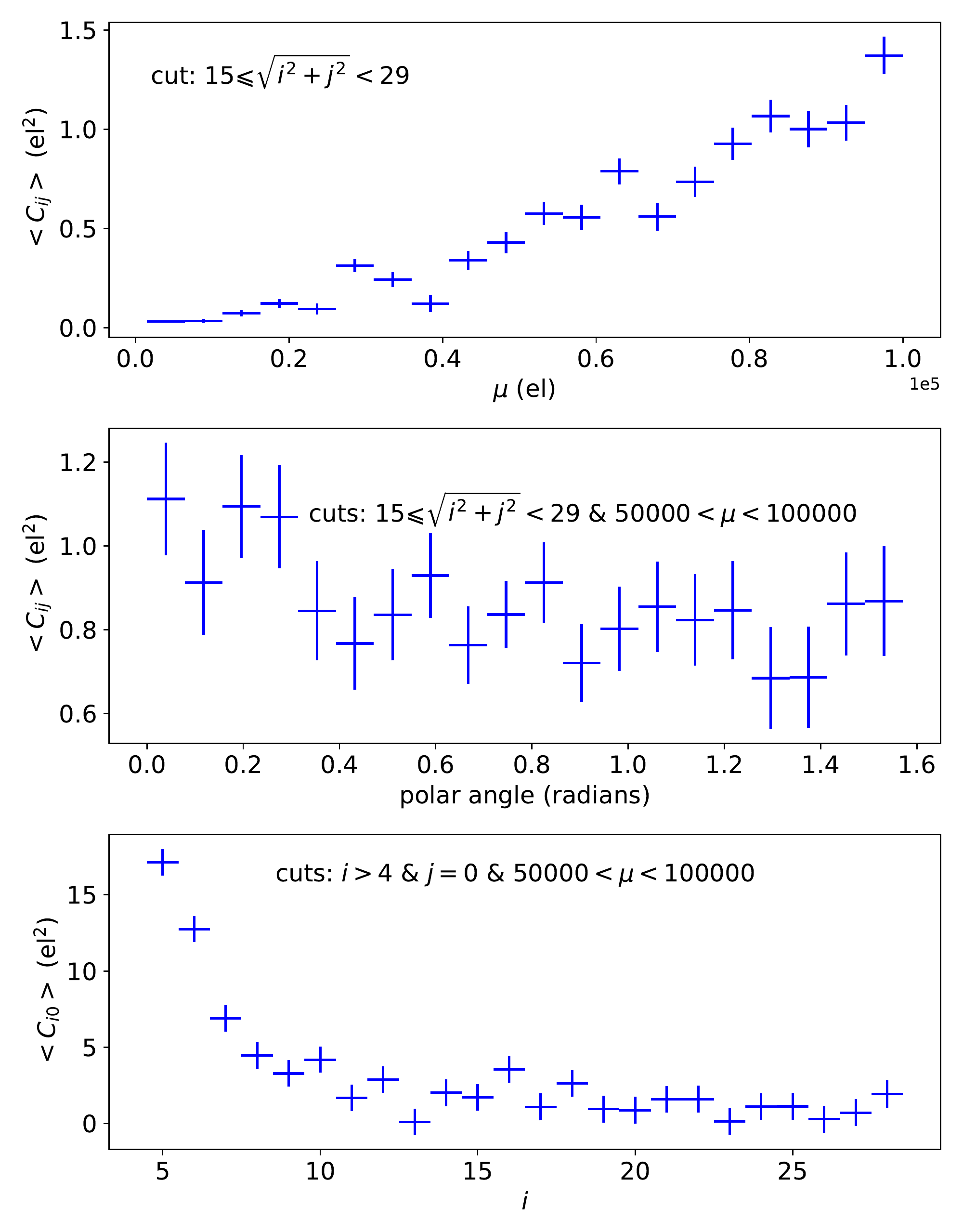}
\caption{These figures display distant covariances, averaged over 
the 16 channels. Top: average value of distant covariances as a function
of flat field average. Middle: the average of covariances 
as a function of angle shows no excess in the serial direction (angle = 0).
Bottom: covariances along the serial direction, which do not exhibit
any obvious structure on top of the decay expected from electrostatics.
\label{fig:distant_cov_plot}}
\end{center}
\end{figure}

In order to identify non-electrostatic contributions to covariances,
we consider covariances at large distances, where electrostatic
contributions are expected to be negligible. From a simple
electrostatic model tuned on the well-measured nearest neighbors
correlations, one can readily infer that electrostatically-induced
covariances beyond a separation of $\sim 12$ are expected to be much
lower than 1~\electron$^2$ at a flat field average of 50 ke, and not
detectable with our statistics. Reproducing the shape of the long
distance decay of electrostatic covariances is not involved because
the source of the disturbing field (i.e. the charge cloud within a pixel
potential well) can then be
considered as a point source without a significant loss of accuracy.

Figure \ref{fig:distant_cov_plot} shows that long-distance covariances
are larger than expected from electrostatics only and they are increasing
with the flat field average. We do not know the cause of this small
distant signal, 
which seems almost isotropic. It is however smaller than
covariances up to separations of about 8, and so does not deserve
a correction for the analysis we did in the previous section.
The bottom plot of figure \ref{fig:distant_cov_plot} shows the
serial correlation decaying as expected, but at some point, it displayed an oscillating
pattern associated to a gain oscillation at 40 kHz induced by a pickup
affecting the clocking of the FPGA and in turn the duration of the analog
integration windows for signal and pedestal. This was solved by damping
the injection by the switching power supply sourcing the noise, and using a
phase-locked loop on the FPGA clocks.

\section{Discussion}
\label{sec:discussion}

We have developed a framework to analyse photon transfer curves and
covariance curves, starting from some model of the effect of stored charges on
incoming currents. This framework allows one to constrain the
electrostatic distortions as a function of separation and charge level,
using large-statistics flat field data, covering a large dynamical range.
In order to exploit our high-statistics data set, we had to correct for the
read out chain non linearity and deferred charge effects. In
a separate study, we are investigating if altering the operational voltages of the sensor
reduces deferred charge effects.

Our model shows that approximating the charge-induced pixel area changes
by the slope of correlation curves can lead to sizable errors, depending
on the sensor, but typically scaling as $a_{00} \mu$, and which in turn bias
the $a_{ij}$ measurements by a few \%. 
Our analysis of flat field data shows that, for the e2v sensor we are
studying (and under the chosen operational conditions),  assuming that
pixel boundaries shift linearly with source charges should be questioned.
We have not yet explored the practical consequences of this finding when
it comes to modeling the brighter-fatter effect on science images.
One should remark that the image contrasts in uniform images are small
  as compared to the ones in science images, by typically two orders of magnitude or more. One should question if the low-constrast measurements carried out in
  flat fields apply directly to higher constrast images (see e.g. \cite{Rasmussen16}). We are then attempting to devise practical tests of this apparent
  non-linearity using high contrast images.

Finally, we note that all attempts to correct the brighter-fatter
effect following the framework proposed in \cite{ Guyonnet15}
typically leave of the order of 10 \% of the initial effect in the images
\citep{Guyonnet15, Gruen-PACCD-15, Mandelbaum-HSC}. We have identified
in this work a set of effects that could contribute to these residual
PSF variation with flux by biasing the measurements of electrostatic
forces at play: non-linearity of the electronic chain, covariances
induced by non-electrostatic sources, oversimplification of the flux
dependence of variance and covariances, small biases in variance and
covariance estimates. These small contributions can conspire to leave
a flux-dependent PSF after correction of the images for the
brighter-fatter effect.

\begin{acknowledgements}
  We are grateful to the anonymous referee for a careful reading of the article and for providing very useful suggestions. This paper has undergone internal review in the LSST Dark Energy Science Collaboration. 
  The internal reviewers were C. Lage, A. Nomerotski and A. Rasmussen, and their remarks significantly improved the manuscript. 
  All authors contributed extensively to the work presented in this paper.
  This research was mostly funded by CNRS (France). 
  The DESC acknowledges ongoing support from the Institut National de Physique Nucl\'eaire et de Physique des Particules in France; the Science \& Technology Facilities Council in the United Kingdom; and the Department of Energy, the National Science Foundation, and the LSST Corporation in the United States.  DESC uses resources of the IN2P3 Computing Center (CC-IN2P3--Lyon/Villeurbanne - France) funded by the Centre National de la Recherche Scientifique; the National Energy Research Scientific Computing Center, a DOE Office of Science User Facility supported by the Office of Science of the U.S.\ Department of Energy under Contract No.\ DE-AC02-05CH11231; STFC DiRAC HPC Facilities, funded by UK BIS National E-infrastructure capital grants; and the UK particle physics grid, supported by the GridPP Collaboration.  This work was performed in part under DOE Contract DE-AC02-76SF00515.

\end{acknowledgements}

\def\aap{A\&A}
\def\nat{Nature}
\def\mnras{MNRAS}
\def\memras{MmRAS}
\def\aapr{The Astron. and Astrop. Rev.}
\def\prd{Phys. Rev. D}
\def\sovast{Soviet Astronomy}
\def\jcap{J. Cosm. Astropart. P.}
\def\aj {Astron. Journ.}
\bibliography{biblio}

\begin{appendix}
  \section{Computation of covariances in the Fourier domain}
  \label{app:comp_fourier}

We provide here a sample of the Python code we use to compute
covariances in the Fourier domain, together with the calculation in
direct space used to check that results are identical, and to compare the
computing speeds. In order to avoid that pixels at the end of (e.g.) a
line are correlated with the ones at the beginning, one has to
zero-pad the data provided as input to the discrete Fourier transform
(DFT), in both directions, with at least as many zeros as the maximum
considered separation. With \verb+numpy.fft+, zero-padding is done internally
by providing the wanted size (called shape in Python parlance) of the
DFT. One technical note is required here: when using DFT for real
images (\verb+numpy.fft.rfft2+), one should provide an even second
dimension for the direct transform in order for the inverse transform
(\verb+irfft2+) to return the original data (up to rounding errors).

Here are the Fourier and direct routines :
\begin{small}
\begin{verbatim}
import numpy as np

def compute_cov(diff, w, fft_shape, maxrange) :
    """
    diff : image to compute the covariance of
    w : weights (0 or 1) of the pixels of diff
    fft_shape : the actual shape of the DFTs
    maxrange: last index of the covariance to be computed
    returns cov[maxrange, maxrange], npix[maxrange,maxrange]
    """
    assert(fft_shape[0]>diff.shape[0]+maxrange)
    assert(fft_shape[1]>diff.shape[1]+maxrange)
    # for some reason related to numpy.fft.rfftn,
    # the second dimension should be even, so
    my_fft_shape = fft_shape
    if fft_shape[1] %2 == 1 :
        my_fft_shape = (fft_shape[0], fft_shape[1]+1)
    # FFT of the image
    tim = np.fft.rfft2(diff*w, my_fft_shape)
    # FFT of the mask
    tmask = np.fft.rfft2(w, my_fft_shape)
    # three inverse transforms:
    pcov = np.fft.irfft2(tim*tim.conjugate())
    pmean= np.fft.irfft2(tim*tmask.conjugate())
    pcount= np.fft.irfft2(tmask*tmask.conjugate())
    # now evaluate covariances and numbers of "active" pixels
    cov = np.ndarray((maxrange,maxrange))
    npix = np.zeros_like(cov)
    for dx in range(maxrange) :
        for dy in range(maxrange) :
            # compensate rounding errors
            npix1 = int(round(pcount[dy,dx]))
            cov1 = pcov[dy,dx]/npix1-pmean[dy,dx]*pmean[-dy,-dx]/(npix1*npix1)
            if (dx == 0 or dy == 0):
                cov[dy,dx] = cov1
                npix[dy,dx] = npix1
                continue
            npix2 = int(round(pcount[-dy,dx]))
            cov2 = pcov[-dy,dx]/npix2-pmean[-dy,dx]*pmean[dy,-dx]/(npix2*npix2)
            cov[dy,dx] = 0.5*(cov1+cov2)
            npix[dy,dx] = npix1+npix2
    return cov,npix

def cov_direct_value(diff ,w, dx,dy):
    (ncols,nrows) = diff.shape
    if dy>=0 :
        im1 = diff[dy:, dx:]
        w1 = w[dy:, dx:]
        im2 = diff[:ncols-dy, :nrows-dx]
        w2=w[:ncols-dy, :nrows-dx]
    else:
        im1 = diff[:ncols+dy, dx:]
        w1 = w[:ncols+dy, dx:]
        im2 = diff[-dy:, :nrows-dx]
        w2 = w[-dy:, :nrows-dx]
    w_all = w1*w2
    npix = w_all.sum()
    im1_times_w = im1*w_all
    s1 = im1_times_w.sum()/npix
    s2 = (im2*w_all).sum()/npix
    p = (im1_times_w*im2).sum()/npix
    cov = p-s1*s2
    return cov,npix
    
\end{verbatim}
\end{small}
\end{appendix}

\end{document}